\documentclass[preprint2]{aastex}

\usepackage{subfigure}
\usepackage{booktabs}
\usepackage{hyperref}
\usepackage{threeparttable}
\usepackage{filecontents}
\usepackage{float}
\usepackage{array}

\shorttitle{Magnetic cycles in the solar-like stars Kepler-17 and Kepler-63}
\shortauthors{Estrela and Valio}

\begin{document}


\title{Stellar magnetic cycles in the solar-like stars Kepler-17 and Kepler-63}


\author{Raissa Estrela and Adriana Valio}

\affil{Center for Radio Astronomy and Astrophysics (CRAAM),
Mackenzie Presbyterian University, Rua da Consolacao 896, Sao Paulo,
SP 01302-907, Brazil}
\email{rlf.estrela@gmail.com}
\email{avalio@craam.mackenzie.br}






\begin{abstract}
The stellar magnetic field plays a crucial role in the star internal mechanisms, as in the interactions with its environment. The study of starspots provides information about the stellar magnetic field, and can characterise the cycle. Moreover, the analysis of solar-type stars is also useful to shed light onto the origin of the solar magnetic field.  The objective of this work is to characterise the magnetic activity of stars. Here, we studied two solar-type stars Kepler-17 and Kepler-63 using two methods  to estimate the magnetic cycle length. The first one characterises the spots (radius, intensity, and location) by fitting the small variations in the light curve of a star caused by the occultation of a spot during a planetary transit. This approach yields the number of spots present in the stellar surface and the flux deficit subtracted from the star by their presence during each transit. The second method estimates the activity from the excess in the residuals of the transit lightcurves. This excess is obtained by subtracting a spotless model transit from the lightcurve, and then integrating all the residuals during the transit. The presence of long term periodicity is estimated in both time series. With the first method, we obtained $P_{\rm cycle}$ = 1.12 $\pm$ 0.16 yr (Kepler-17) and $P_{\rm cycle}$ = 1.27 $\pm$ 0.16 yr (Kepler-63), and for the second approach the values are 1.35 $\pm$ 0.27 yr and 1.27 $\pm$ 0.12 yr, respectively. The results of both methods agree with each other and confirm their robustness.

\end{abstract}  


\keywords{magnetic activity, magnetic cycles, starspots}



\section{Introduction}

Magnetic fields comparable to that of our Sun appear in the internal regions of all cool, low-mass solar-type stars, in small or higher scales \citep{lammer15}. The presence of the magnetic field can affect the evolutionary stages of a star and is responsible for its magnetic activity. 
The evidence of the magnetic field presence is the emergence of dark spots in active regions on the stellar surface. The number of spots appearing and disappearing in the stellar disk varies in cycles. This behaviour observed in the Sun shows an 11 year-cycle, but it also happens in other stars.
 
The HK-project using the Mount Wilson Observatory, investigated hundred of stars by observing the Ca II HK lines to find stellar cycles similar to the solar case, with 52 of them showing this periodic behaviour \citep{bau95}. Using the Mount Wilson data, \cite{saar99} also studied stellar cycles and multiple cycles. From these observations, they estabilished the relation between stellar rotation period, $P_{\rm rot}$, and magnetic cycle period, $P_{\rm cycle}$ as a function of the Rossby number, dividing the stars into active (A) and inactive (I) branches, which were distinguished by the number of rotations per cycle and activity level. Later, several long photometric records for a significant sample of stars became available, and other studies of the stellar cycles were performed \citep{olah09, mg02, lovis11}. 

Along with the 11 year cycle of the Sun, there are other short-durations cycles, known as quasi-biennial oscilations (QBO). \cite{fletch10} found a quasi-biennial period of 2 years in the low-degree solar oscillation frequencies of the Sun after separating this signal from the influence of the dominant 11 year solar cycle. In addition, \cite{bazil14} reported  solar QBOs with the time scale of 0.6$-$4 years. In other stars, some short period variations were also reported. The first were detected by \cite{bau95} in two stars from the Mount Wilson survey, HD 76151 with 2.52 years and HD 190406 with 2.60 years. \cite{olah09} found fifteen stars showing multiple cycles by the analysis of Ca II emission. Using high cadence SMARTS HK data, \cite{meta10} discovered a 1.6 year magnetic activity cycle for the solar twin $\iota$ Horologii and found two cycles in $\epsilon$ Eridani, one of 2.95 years and a long-term cycle of 12.7 years. \cite{ege15} also found a short-period $\sim$ 1.7 year for the solar analogue HD 30495. Using Kepler observations,  \cite{salabert16} discovered from the analysis of frequency shifts a 1.5 year variation for the solar analog KIC 10644253, that could be a short-period or quasi-biennal oscillation.


Other techniques such as monitoring starspots can provide information on stellar cycles. Evidence of the spots can be found by analysing planetary transits in the star lightcurve \citep{nutz11,sanchis11,soap13}. Small variations in the lightcurve occur because when the planet eclipses the star it may occult spots on the stellar surface, causing an increase in the luminosity during the transits. \cite{sil03} developed  a method to infer the properties of starspots from high precision transit photometry, such as size, position and intensity. Here we use this model to characterize the magnetic cycle of two active solar-type stars: Kepler-17 and Kepler-63. To cross check this model, we adopted a new technique, called here as transit residuals excess, to verify the variation of the level of activity in both stars.

This paper is organized as follows. Section \ref{sec:observations} provides an overview of the observations for Kepler-17 and Kepler-63. Section \ref{sec:spotmodel} explains the first method adopted in this work, whereas the next section describes the physical parameters of the modelled spots. Section \ref{sec:starmag} presents the second method and the results for the activity cycles obtained with both methods. The paper finishes with the discussions and conclusions, presented in Sections \ref{sec:disc} and \ref{sec:conclu}, respectively.

\section{Observations}
\label{sec:observations}

The stars analysed in the present work, Kepler-17 and Kepler-63, were observed by the Kepler satellite. This mission was responsible for collecting data of thousand of stars and planets (\cite{boru10}). The duration of the mission was scheduled for 3.5 years, initially planned to finish in 2012 but was extended to 2016 (K2 mission) (see \cite{howell14}). The telescope has an aperture of 95 cm and explores about $1.6 \times 10^{5}$ stars in a field of 150deg$^{2}$. It was projected to discover Earth size planets located in the habitable zone of stars (dwarfs F to K). To keep the solar panel towards the Sun, the spacecraft needed to rotate about its axis by 90$\,^{\circ}$ every 93 days, this time interval is known as ``quarter''. A total of 2,327 planets were confirmed until now and there are 4706 planet candidates \footnote{http://kepler.nasa.gov/, July 2017}.

The lightcurves can be obtained from the MAST \footnote{http://archive.stci.edu/} (\textit{Mikulski Archive for Space Telescopes}) database. The observation of the target stars are made along sixteen quarters and consists of long (one data point each 29.4 minutes) and short cadence data ($\sim$ 1 min time resolution). For our study, we selected only short cadence lightcurves, because we are interested in short time scale variations, with planetary transits over the whole period of operation of the satellite. 

The stars analysed here are active solar type stars. Kepler-17 is a G2V star with age $<$ 1.8 Gyr and Kepler-63 is also a young G-type star (subclass still unknown) of about 200 Myr. Both stars are accompanied by a giant planet in close orbit (Hot Jupiter): Kepler-17b has $\approx 2.5$ Jupiter masses and an orbital period of 1.486 day (\cite{desert11}), while Kepler-63b has $\approx$ 0.4 Jupiter masses and an orbital period of 9.434 days. During the 1240 days of observation of the Kepler mission, a total 834 possible transits were detected for Kepler-17, while Kepler-63 was observed for 1400 days and 150 transits were registered. The lightcurve of these stars shows that they are very active and marked by rotational modulations caused by the presence of starspots, with a peak-to-peak variation of $6\%$ for Kepler-63 and $4\%$ for Kepler-17. The active regions of Kepler-17 was previously analysed by \cite{bonomo12}, who found spots area of 10 to 15 times larger than the solar ones and evidences of a differential rotation in latitudes similar to the those presented in the Sun. Study of spot evolution in this star was also perfomed by \cite{davenport15}. The physical properties of the stars and their planets are shown in Table \ref{table:tab1} and Table \ref{table:tab2}, respectively.

\begin{table*}[!ht]
\refstepcounter{table}\label{table:tab1}
\resizebox{0.8\textwidth}{!}{\begin{minipage}{\textwidth}
\centering
\begin{threeparttable}
\begin{tabular}{cccccccc}
\multicolumn{8}{c}{\textbf{Table 1}}                                                                                                                                     \\
\multicolumn{8}{c}{Observational parameters of the stars}                                                                                                                \\
\toprule
\toprule
           & KIC Number & Mass {[}$M_{\odot}${]}   & Radius {[}R$_{\odot}${]}   & Age {[}Gyr{]}    & Effective temperature {[}K{]} & Rotation period {[}d{]} & Reference \\
Kepler-17 & 10619192   & 1.16 $\pm$ 0.06          & 1.05 $\pm$ 0.03            & $<$ 1.78 Gyr     & 5780 $\pm$ 80                 & 11.89 $\pm$ 0.15        & 1,2       \\
Kepler-63 & 11554435   & 0.984$^{+0.04}_{-0.035}$ & 0.901$^{+ 0.022}_{-0.027}$ & 0.2 Gyr & 5580 $\pm$ 50                 & 5.401 $\pm$ 0.014       & 3        \\
\bottomrule
\end{tabular}
\begin{tablenotes}
\item \textbf{References.} (1) \cite{bon12}, (2) \cite{desert11} and (3) \cite{sanchis13}.
\end{tablenotes}
\end{threeparttable}
\end{minipage}}
\end{table*}

\begin{table*}[!ht]
\refstepcounter{table}\label{table:tab2}
\resizebox{0.7\textwidth}{!}{\begin{minipage}{\textwidth}
\centering
\begin{threeparttable}
\begin{tabular}{llllllllll}
\multicolumn{10}{c}{\textbf{Table 2}}                                                                                                                                                                                                                                                                                    \\
\multicolumn{10}{c}{Observational parameters of the planets}                                                                                                                                                                                                                                                             \\
\toprule
\toprule
          & KIC Number                     & Mass {[}$M_{\rm jup}${]}                 & Radius {[}$R_{\rm jup}${]}               & Radius {[}$R_{\rm star}${]}          & Orbital                   & Inclination, $i$ {[}deg{]}                    & Semi-major                       & Semi-major                      & Reference \\
          &                                &                                      &                                      &                                  & Period {[}d{]}            &                                               & axis, $a$ {[}AU{]}               & axis, $a$ {[}$R_{\rm star}${]}      &           \\
Kepler-17b & \multicolumn{1}{c}{10619192 b} & \multicolumn{1}{c}{2.45 $\pm$ 0.014} & \multicolumn{1}{c}{1.442$^{a}$}      & \multicolumn{1}{c}{0.138$^{a}$}  & \multicolumn{1}{c}{1.485} & \multicolumn{1}{c}{87.2 $\pm$ 0.15}           & \multicolumn{1}{c}{0.0279$^{a}$} & \multicolumn{1}{c}{5.729$^{a}$} & 1,2       \\
Kepler-63b & \multicolumn{1}{c}{11554435 b} & \multicolumn{1}{c}{$<$ 0.3775}       & \multicolumn{1}{c}{0.593$^{a}$} & \multicolumn{1}{c}{0.0662$^{a}$} & \multicolumn{1}{c}{9.434} & \multicolumn{1}{c}{87.8 $^{+0.018}_{-0.019}$} & \multicolumn{1}{c}{0.081$^{a}$}  & \multicolumn{1}{c}{19.35$^{a}$} & 3     \\
\bottomrule   
\end{tabular}
\begin{tablenotes}
\item $^{a}$ Modified values to obtain a better fit of each transit lightcurve.
\item \textbf{References.} (1) \cite{bon12}, (2) \cite{desert11} and (3) \cite{sanchis13}.
\end{tablenotes}
\end{threeparttable}
\end{minipage}}
\end{table*}





\section{Spot modelling}
\label{sec:spotmodel}

\begin{figure}[ht]
  \centering
\includegraphics[scale=0.40]{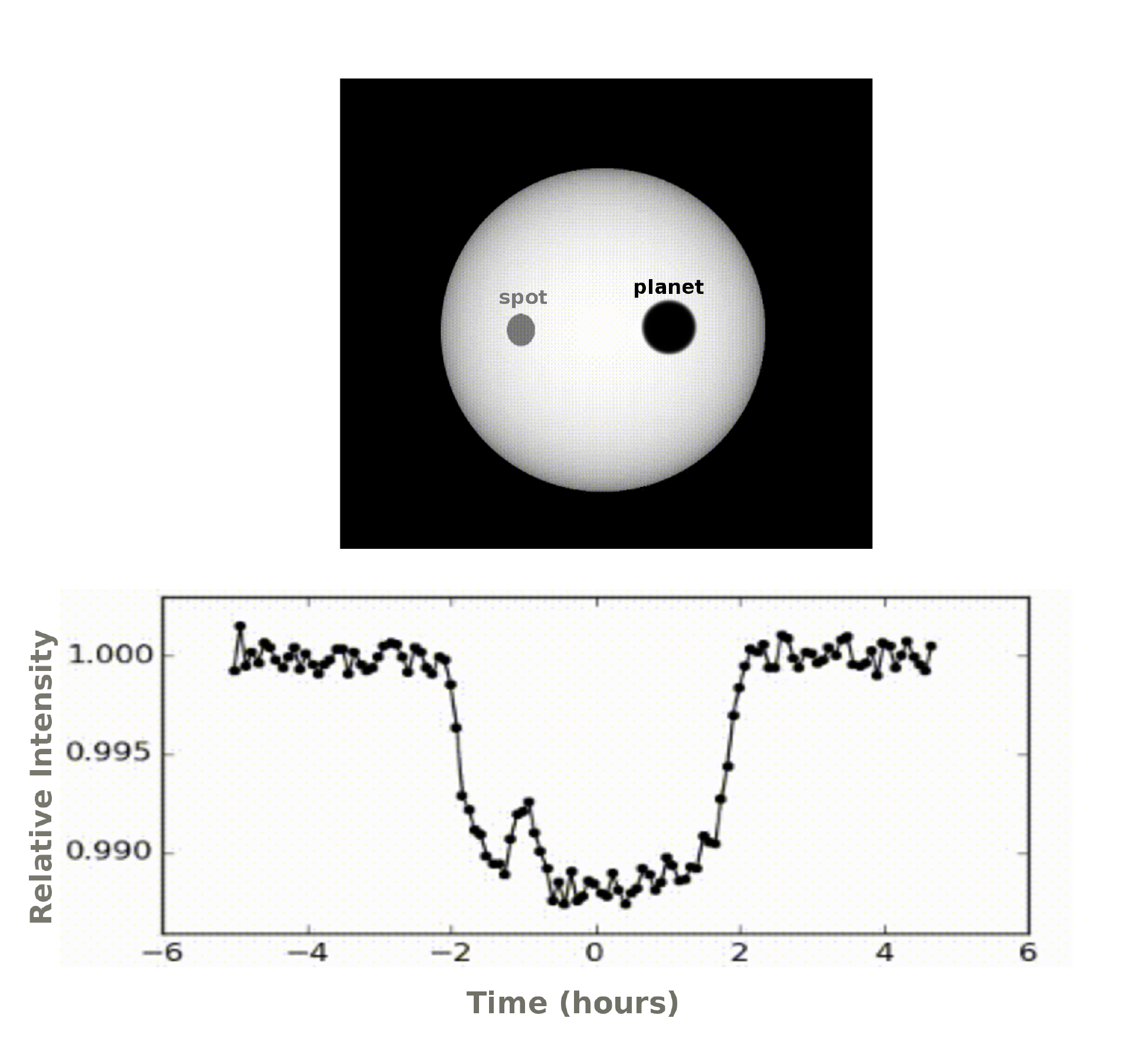}
  \caption[Model]{\textit{Top}: 2D Simulated image of a star with quadratic limb darkening and one spot, and its planet, assumed as a dark disk. \textit{Bottom}: Resulting light curve from the transit, with the ``bump'' due to spot-crossing by the planet.} 
\label{fig:figura1_1}
\end{figure}

To characterize the spots, we adopted the transit model proposed by \cite{sil03} that uses small transit photometry variations to infer the properties of starspots. The passage of the planet in front of the star may occult solar-like spots on the stellar surface, producing a slight increase in the luminosity detected during a short period (minutes) of the transit. This occurs because the spot is darker (cooler region) than the stellar photosphere and causes a smaller decrease in the intensity than when it blocks a region without spots. This effect is shown in Figure ~\ref{fig:figura1_1}.

This model simulates a star with quadratic limb darkening as a 2D image and the planet is assumed to be a dark disk with radius $R_{p}/R_{\rm star}$, where $R_p$ is the radius of the planet and $R_{\rm star}$ is the radius of the primary star. For each time interval, the position of the planet in its orbit is calculated according to its parameters: inclination angle $i$ and semi-major axis $a$. The simulated lightcurve results from the intensity integration of all the pixels in the image for each planet position during the transit. All the simulations are performed considering a circular orbit, that is null eccentricity. Applications of such model are described in \cite{sil03} for HD 209458, \cite{sil10} and \cite{sil11} for the active star CoRoT-2.

\begin{figure*}[ht]
  \centering
\includegraphics[scale=0.70]{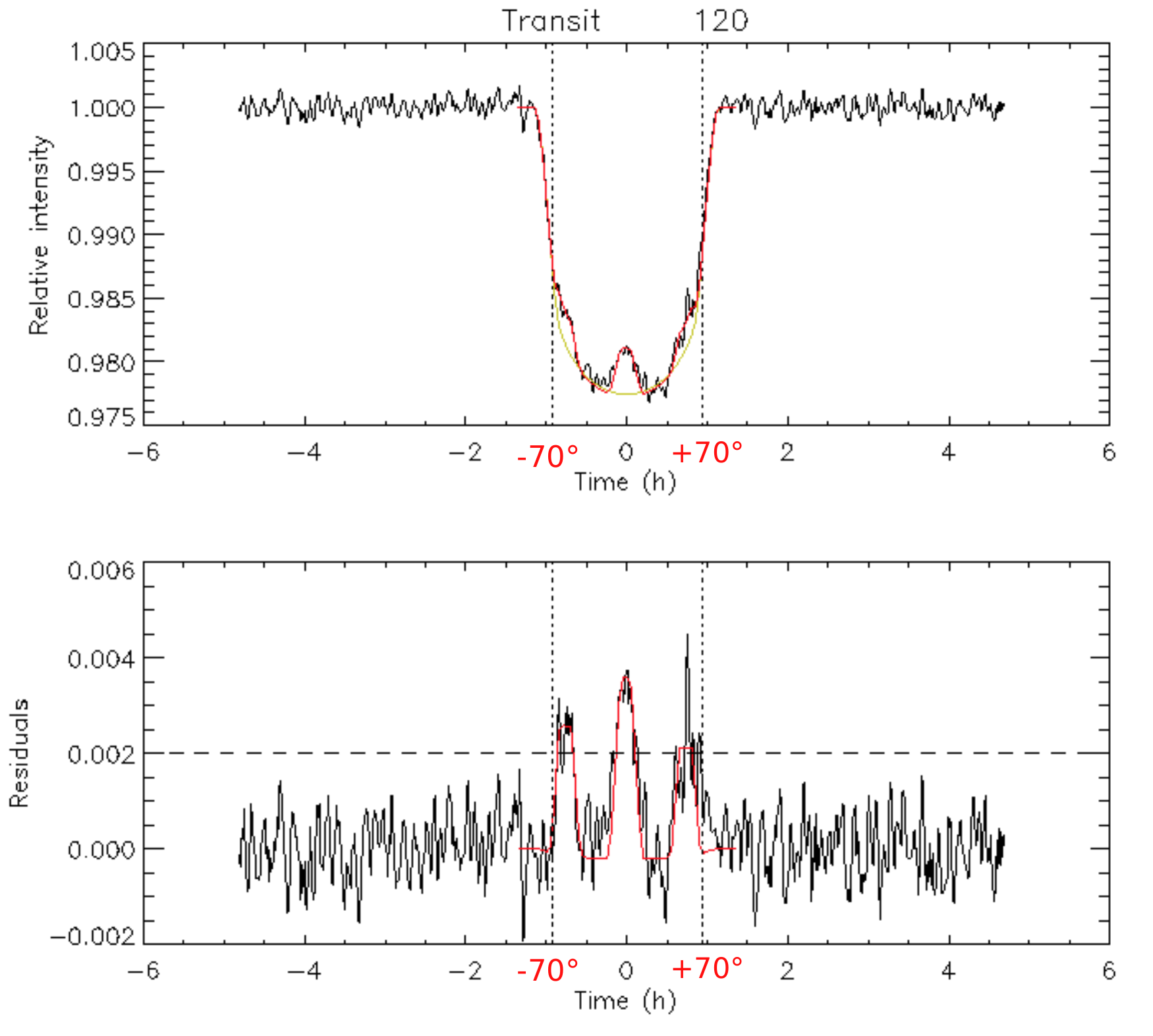}
  \caption[Model]{The 120th transit from Kepler-17 illustrates a typical example of the spot fit by the model developed by \cite{sil03}. \textit{Top}: Transit lightcurve with the model of a spotless star overplotted (yellow). \textit{Bottom}: Residuals of the transit lightcurve after subtraction of a spotless star model. The red curve shows the fit to the data ``bumps'' on both panels.} 
\label{fig:figura1_2}
\end{figure*}

An example of the application of this model is shown in Fig. \ref{fig:figura1_2} for the 120th transit of Kepler-17, where the top panel shows the fit with three spots (red curve), together with the model of a star without any spots (yellow curve). The bottom panel of Fig. \ref{fig:figura1_2}, shows the residuals from the subtraction of the spotless model, where the spots became more evident, as the three ``bumps'' seen in the residuals. An estimation of the noise presented in the Kepler data is given by the CDPP (Combined Differential Photometric Precision, see \cite{christiansen12}), computed for each quarter in the lightcurve. We considered the uncertainty in the data, $\sigma$, as being the average of the CDPP values in all quarters. Only the ``bumps'' that exceed the detection limit of 10$\sigma$ are assumed as spots and modelled. The modelled spots are constrained within longitude of $\pm$ 70$\,^{\circ}$ (dashed lines) to avoid any distortions caused by the ingress and egress branches of the transit. In this region the lightcurve measurements are much less accurate than in the central part of the transit due to the steep gradients in intensity. The maximum number of spots modelled per transit was set to 4, with an exception for only one transit of Kepler-17, where it was necessary to fit 5 spots.

The model assumes that the spots are circular and described by 3 parameters: size (in units of planetary radius $R_p$), intensity with respect to the surface of the star (in units the maximum intensity at disc centre, $I_c$), and  longitude (allowed range is $\pm$70$^{\circ}$ to avoid distortions of the ingress and egress of the transit). For Kepler-17, where the orbital plane is aligned with the stellar equator, the latitude of the spots remains fixed and equal to the planetary transit projection onto the surface of the star. This latitude is -14$\,^{\circ}$.6 and corresponds to an inclination angle of 87.84$\,^{\circ}$, that was arbitrary chosen to be South. Moreover, the foreshortening effect is taken into account for the spots near the limb. Due to the high obliquity of the orbit of Kepler-63b, the planet occults several latitudes of the star from its equator all the way to the poles.

It was necessary to refine the values from the semi-major axis and planet radius taken from the literature \citep{desert11} for Kepler-17b and \citep{sanchis13} for Kepler-63b, to obtain a better fit for each transit lightcurve. Therefore, we used 5.729 $R_{\rm star}$ (Kepler-17b) and 19.35 $R_{\rm star}$ (Kepler-63b) for the semi-major axis, while for the planet radius we adopted 0.0662 $R_{\rm star}$ and 0.138 $R_{\rm star}$, respectively (see Valio et al. 2016 (to be submitted) and Netto and Valio, 2016 (submitted). These values represent 8$\%$ and 6$\%$ (radius) and 4$\%$ and 1$\%$ (semi-major axis) increase, respectively for Kepler-17 and Kepler-63, with respect to the initial values given in \citep{bon12} and \cite{sanchis13}.  


The residuals from each transit lightcurve were fit using this model. We performed initial guesses manually, for the longitude of the spot, $lg_{spot}$, obtained from the approximate central time of the ``bump'', $t_{s}$ (in hours), computed as follows:

\begin{equation}
t_{s} = \left\lbrace 90\,^{\circ} - \arccos \left[ \frac{\sin(lg_{s})) \cos(lat_{s})}{a} \right] \right\rbrace \frac{P_{\rm orb}}{360 ^{\circ}} 24 \rm h
\end{equation}

\begin{figure}[!ht]
  \centering
\includegraphics[scale=0.18]{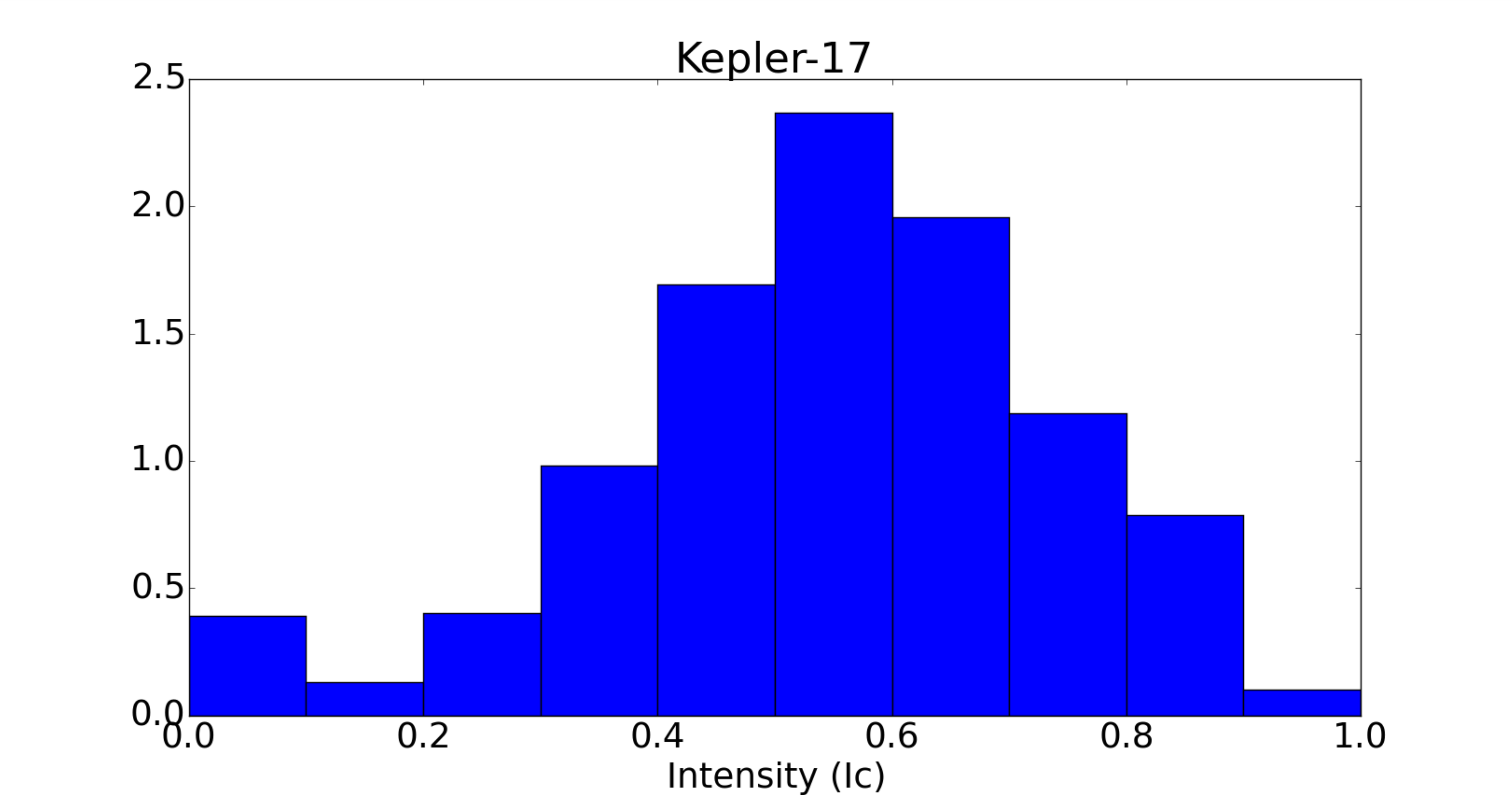}
\includegraphics[scale=0.18]{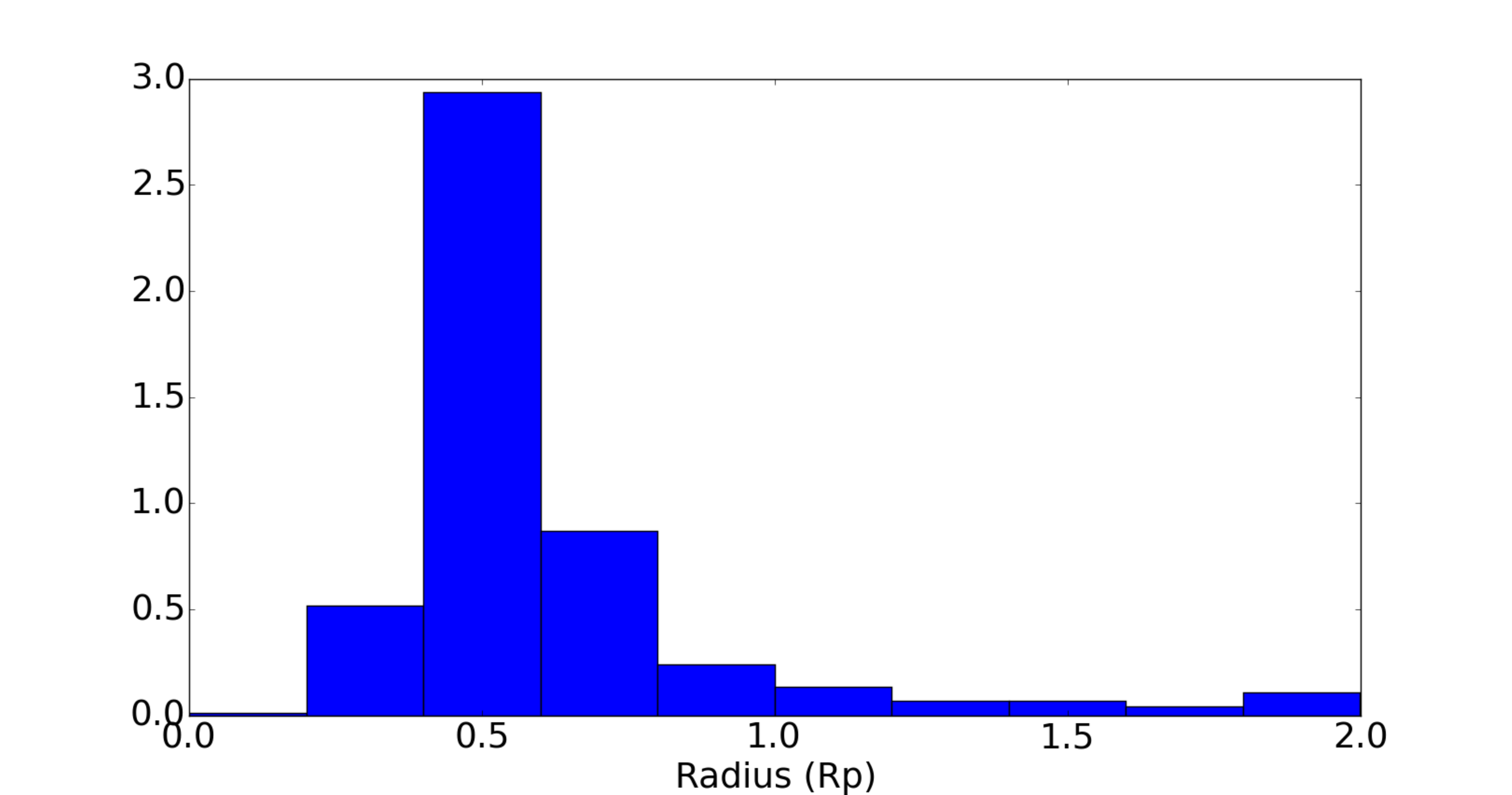}
\includegraphics[scale=0.18]{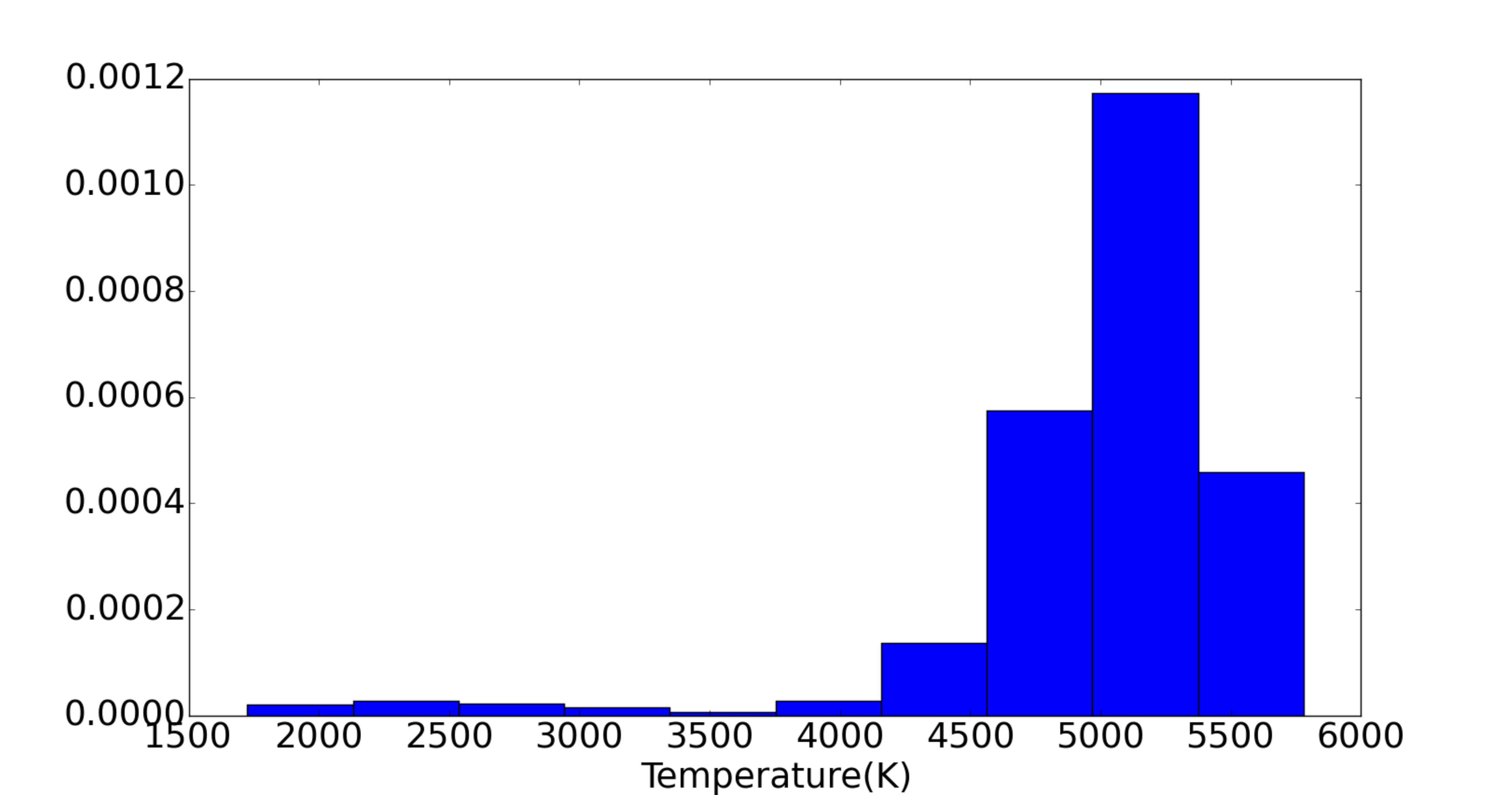}
\includegraphics[scale=0.18]{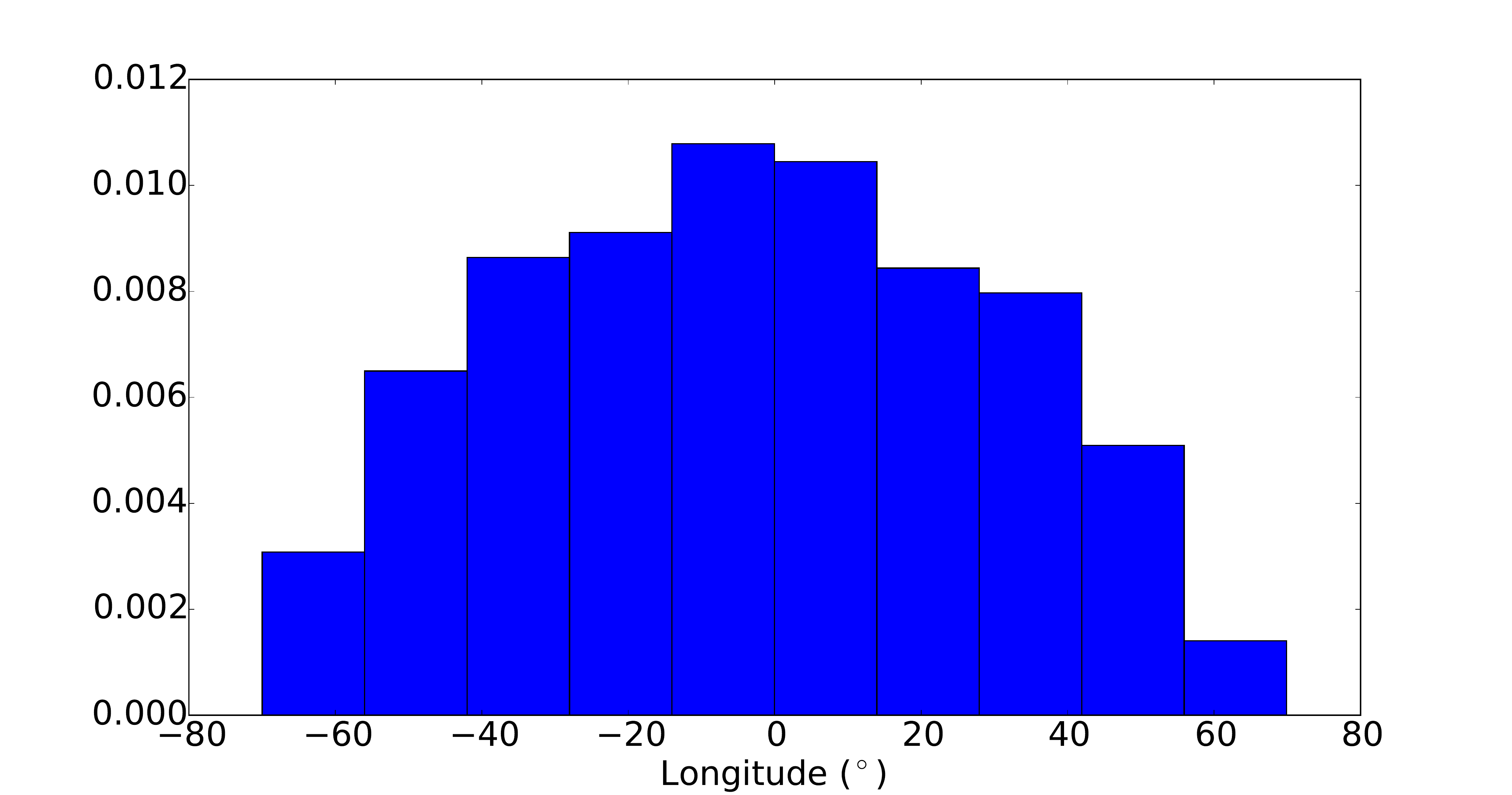}
  \caption[Hist]{Histograms of the spots parameters for Kepler-17: spots intensity (in units of $I_c$) (\textit{top}), spots radius (in units of $R_{p}$) (\textit{middle}), and temperature (\textit{bottom}). } 
\label{fig:figura1_3} 
\end{figure}

\begin{figure}[!ht]
  \centering
\includegraphics[scale=0.18]{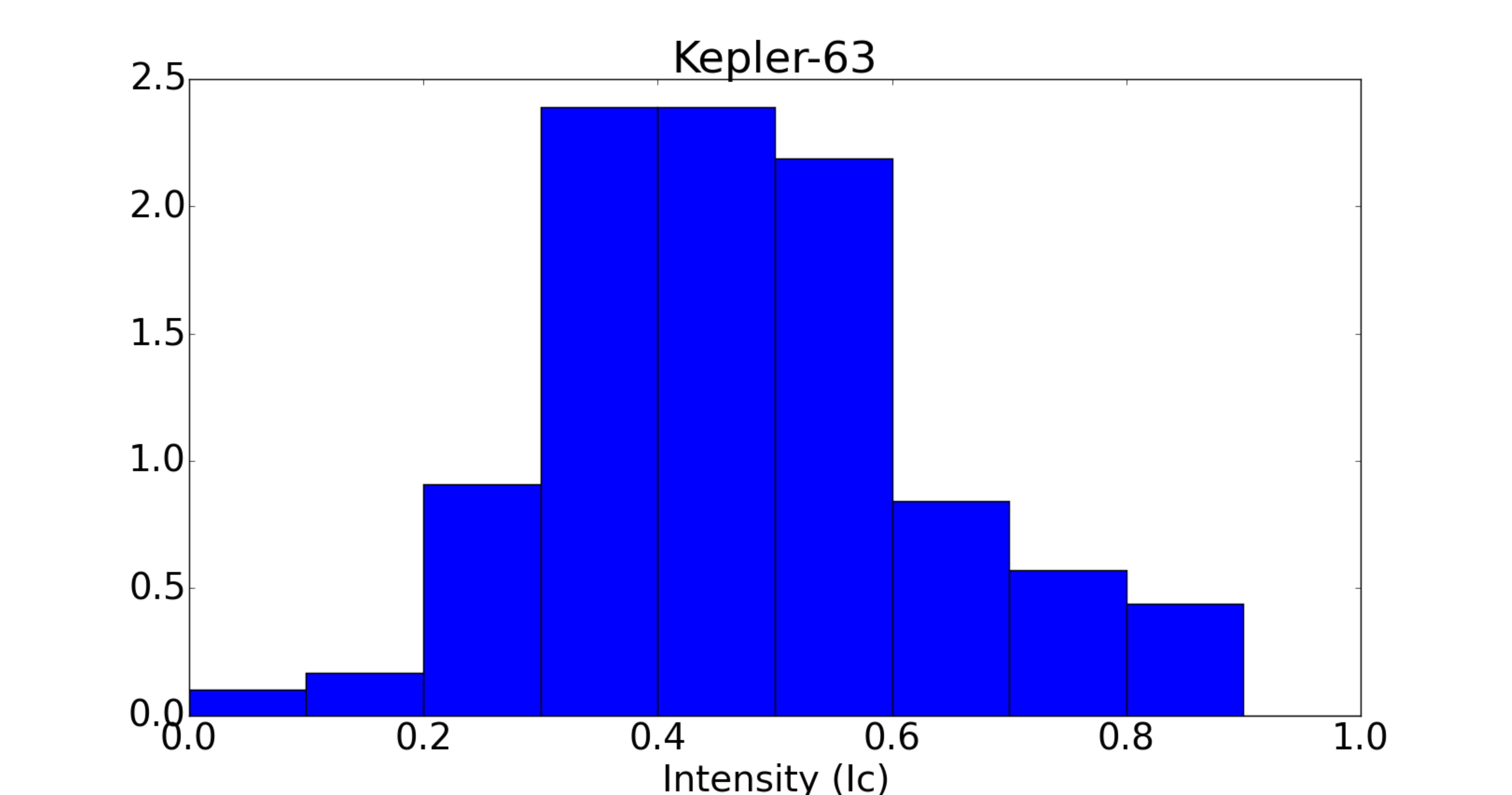}
\includegraphics[scale=0.18]{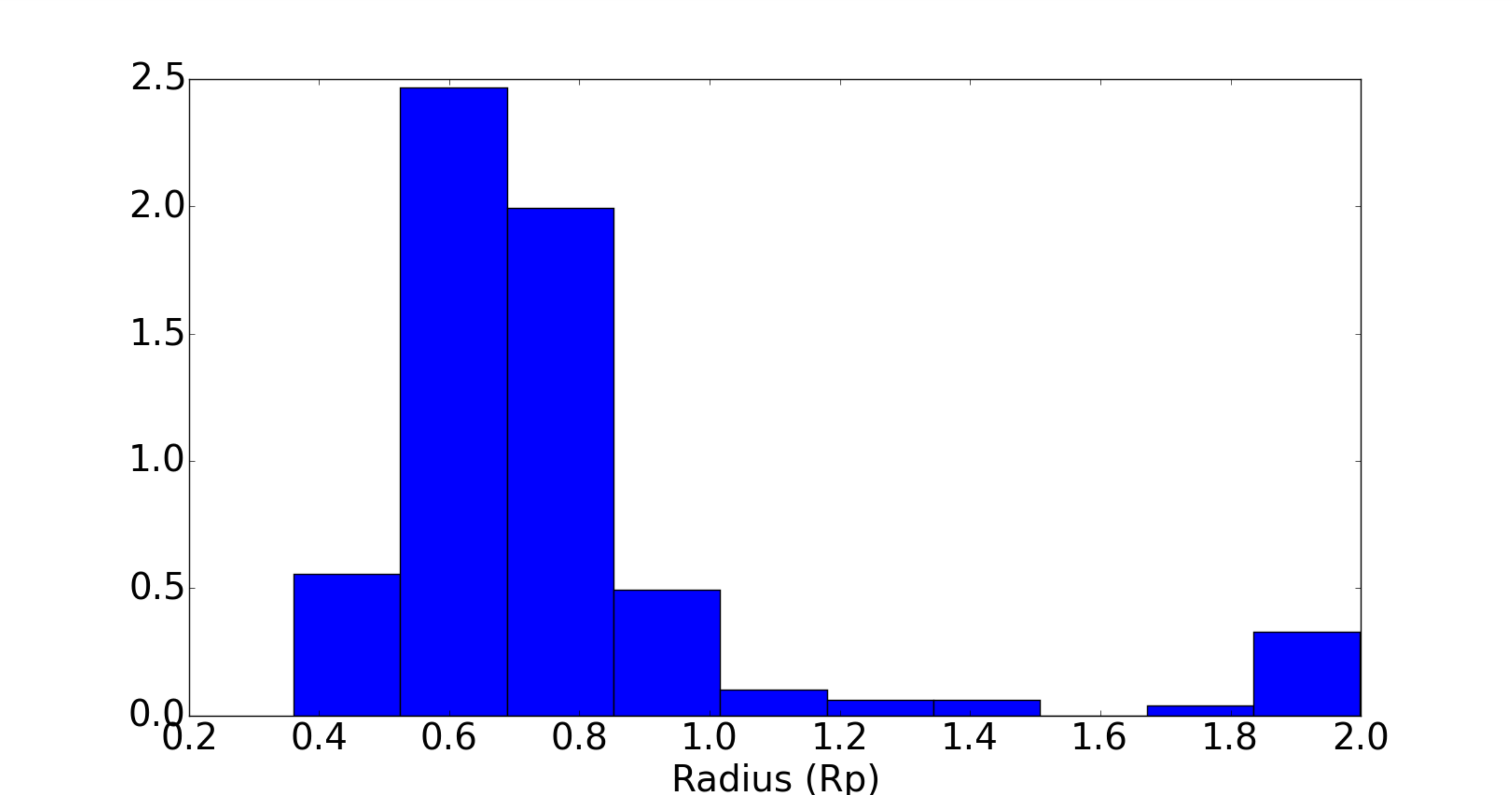}
\includegraphics[scale=0.18]{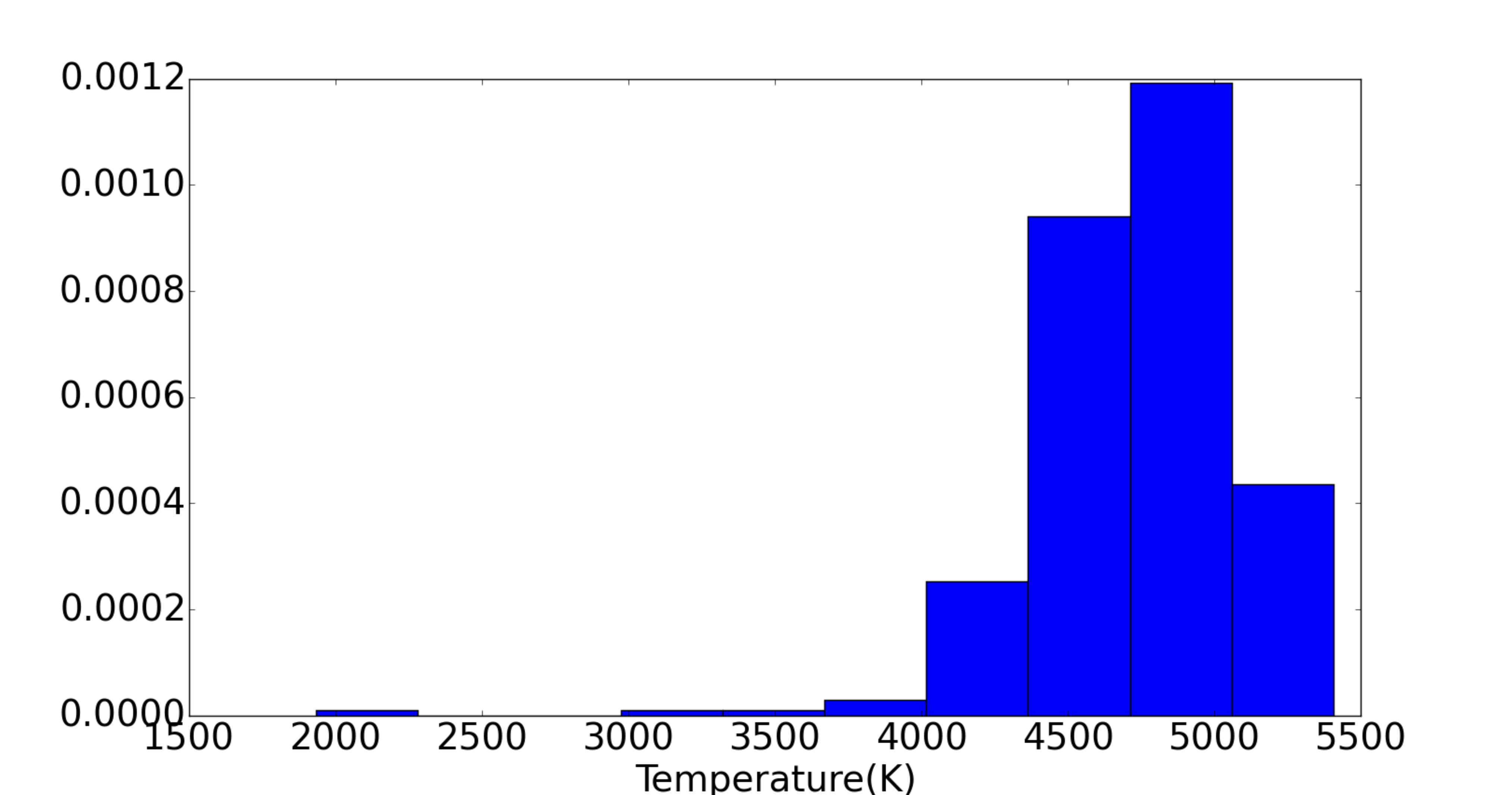}
\includegraphics[scale=0.18]{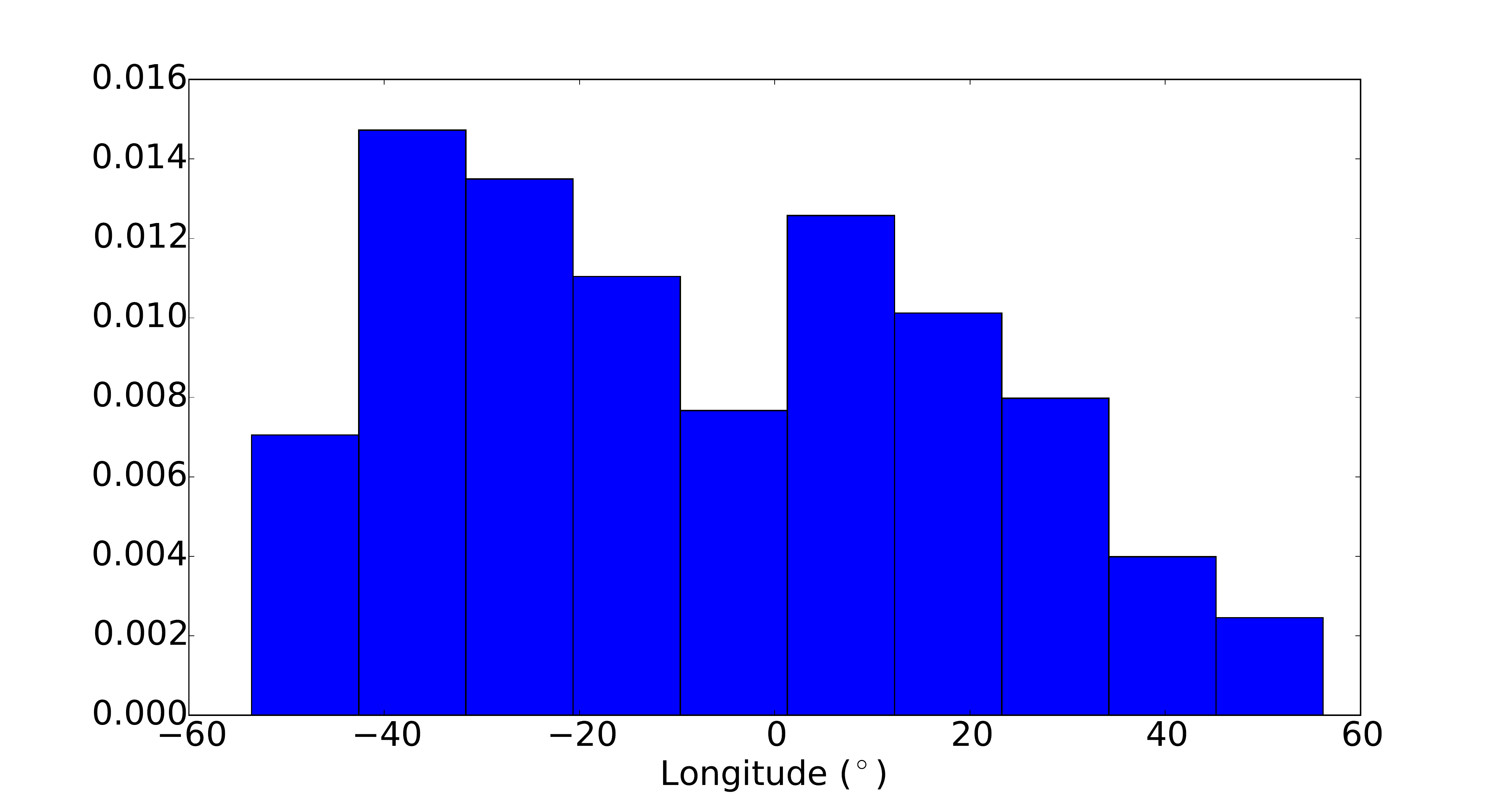}
  \caption[Hist]{Same as Fig. \ref{fig:figura1_3} for Kepler-63.} 
\label{fig:figura1_4}
\end{figure}

\noindent where $P_{\rm orb}$ is the orbital period, $lat_{s}$ is the latitude of the spot in the stellar surface and $a$ is the semi major axis. The number of spots were determined a priori for each transit. For the radius and intensity, we considered initial guesses of 0.5 $R_{p}$ and 0.5 $I_{c}$. The parameters for the spots are chosen from the best fit obtained by the minimization of the $\chi^2$, calculated using the AMOEBA routine \citep{press92}.

\section{Spots physical parameters}
\label{sec:spotpar}

Using the model transit lightcurve proposed by \cite{sil03} and described in the previous section, we analysed each transit separately and the small variations in the luminosity detected during the transit were interpreted as the presence of a spot occulted by the planet. A total of 507 transits for Kepler-17 and 122 for Kepler-63 showed ``bumps'' in the lightcurve residuals above the adopted threshold of 10 CDPP.
From the analysis of these ``bumps'', we obtained a total of 1069 spots (Kepler-17) and 297 (Kepler-63) and estimated their average parameters (radius, intensity and longitude). Furthermore, from the relative spot intensity it is possible to estimate its temperature by considering that both the photosphere and the spots radiate like a blackbody, according to the equation:

\begin{equation}
\frac{I_{\rm spot}}{I_{\rm star}}=\frac{e^{h \nu / K_{B}T_{\rm eff}}-1}{e^{h \nu / K_{B}T_{0}}-1}
\end{equation}
 
\noindent from which we obtain the temperature of the spots:

\begin{equation}
T_{0}= \frac{K_{b}}{h \nu\ \ln \left( f_{i} \left( e^{h \nu/K T_{\rm eff}} -1 \right) +1 \right) }
\end{equation}

\noindent where $K_{b}$ and $h$ are the Boltzmann and Planck constants, respectively, and $\nu$ is the frequency associated with a wavelength of 600 $nm$, $f_{i}$ is the fraction of spot intensity with respect to the central stellar intensity $I_{c}$ obtained from the fit, and $T_{eff}$ is the effective temperature of the star. Considering an effective temperature of 5781 $K$ for Kepler-17 and 5576 $K$ for Kepler-63, we found an average spot temperature of 5000 $\pm$ 600 $K$ and 4800 $\pm$ 400 $K$, respectively, for both stars.

These results are summarized in Table \ref{table:tab3}. The distributions of the parameters are presented in Figure~\ref{fig:figura1_3} and Figure~\ref{fig:figura1_4}. In the case of Kepler-63, we can observe that the radii of the spots are between the range 0.3 up to 2.0 $R_p$. For the intensity, we have a variation from 0.1 to 0.8 $I_c$. The temperature varies from 3000K to 5400K. The longitude varies from -50 to +50$^{\circ}$. For the other star, Kepler-17, the variation of the spots parameters are the following: 0.1 to 2.0 $R_p$ for the radius, 0.1 to 1.0 $I_c$ for intensity, 2000K to 6000K for the temperature, and $\pm$50$^{\circ}$ is the range for the longitude. Note that in Kepler-17 the longitude of the spots are concentrated preferentially in the center of the transit, and for this reason, there is a decrease when the values approach $\pm$70$^{\circ}$, however the same does not happen for Kepler-63, due to the fact that the orbit of the planet crosses several latitudes. For further details of the spot modelling, see Valio et al. (2016) (to be submitted) and Netto $\&$ Valio (2016) (submitted) for Kepler-63.

\begin{table}[!ht]
\centering
\refstepcounter{table}\label{table:tab3}
\begin{tabular}{lcc}
\multicolumn{3}{c}{\textbf{Table 3}}                               \\
\multicolumn{3}{c}{Average values for the parameters of the spots} \\
\midrule
\midrule
                   & Kepler-17                  & Kepler-63               \\
Radius [$R_p$]     & 0.6 $\pm$ 0.3              & 0.8 $\pm$ 0.3   \\
Radius [km]        & (57 $\pm$ 28) $\times 10^{3}$ & (33 $\pm$ 12) $\times 10^{3}$     \\
Intensity [$I_c$]  & 0.54 $\pm$ 0.19            & 0.47 $\pm$ 0.16   \\
Temperature [K]    & 5000 $\pm$ 600             & 4800 $\pm$ 400        \\
Longitude [$^{\circ}$] & -3 $\pm$ 30               & -7 $\pm$ 30 \\
\bottomrule
\end{tabular}
\end{table}

\section{Stellar magnetic activity}
\label{sec:starmag}

Our aim is to characterize the magnetic cycles of Kepler-17 and Kepler-63. For this purpose we adopted two different methods as proxies for the stellar activity: (1) number and flux deficit of the spots modelled in the previous section and (2) estimate of the excess residuals during transits. These procedures were applied to a total of 583 transits for Kepler-17 and 131 transits for Kepler-63. Details and results from both methods are described below.  

\subsection{Spot modelling}
\label{ssec:spotmodel2}

In the case of the Sun, the number of spots on the solar surface is the most common proxy of its activity cycle and displays a periodic variation every 11 years or so. Similarly, the number of spots that appears at the surface of a star varies in accordance with the stellar magnetic cycle. Thus by monitoring the number of spots, during the approximate 4 years of observation of the Kepler stars, it is in principle possible to detect stellar cycles.

\begin{figure}[!ht]
  \centering
\hspace*{-1cm}
\includegraphics[scale=0.25]{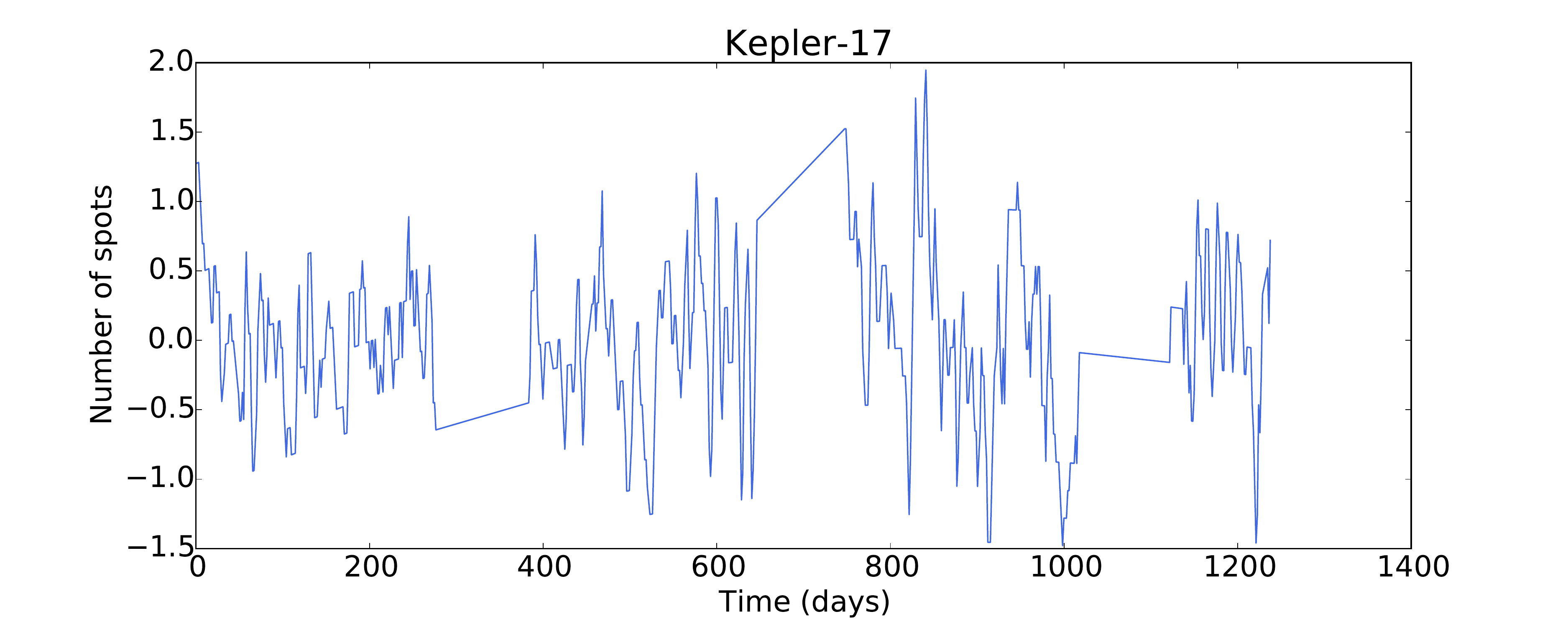}
\hspace*{-1cm}
\includegraphics[scale=0.25]{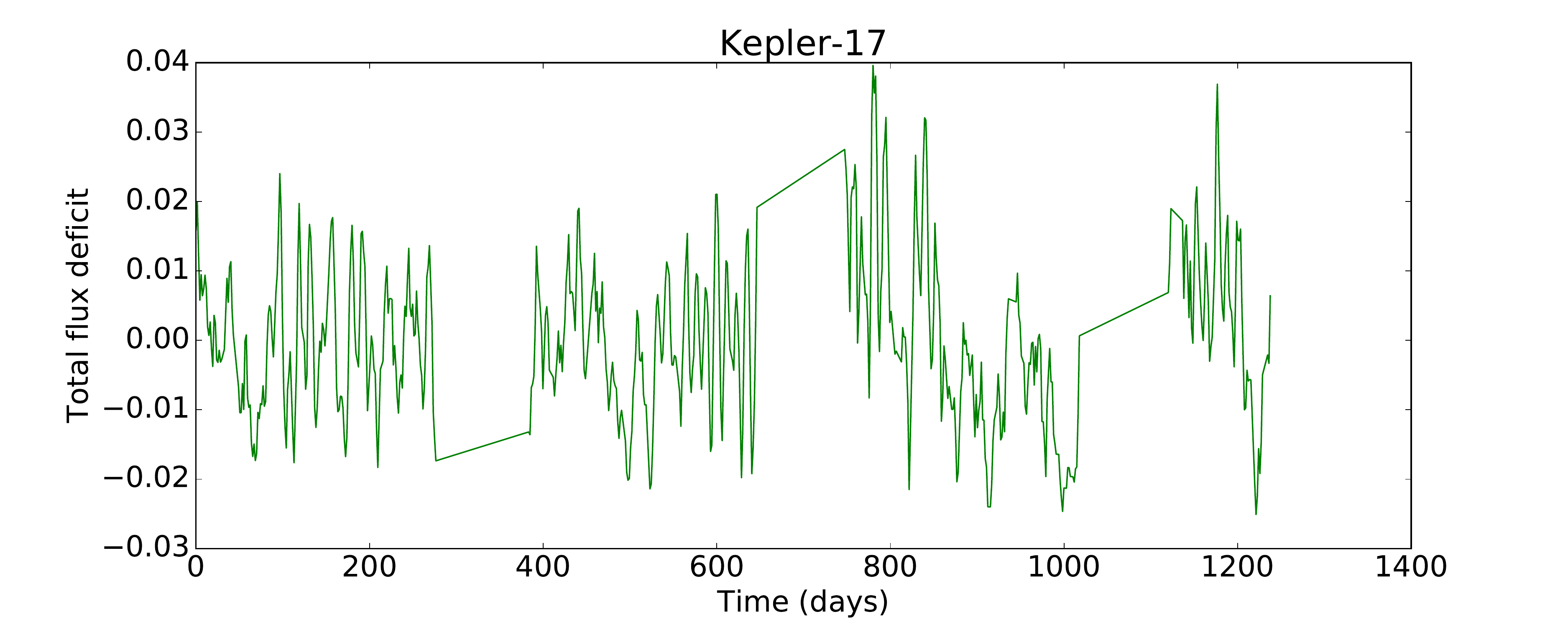}
  \caption[Number]{Number of spots (\textit{Top}) and total flux deficit (\textit{bottom}) caused by the spots in Kepler-17 derived from the transits analysed with the method described in Sections~\ref{sec:spotmodel} and \ref{sec:spotpar}.}
\label{fig:figura1_5}
\end{figure}

\begin{figure}[!ht]
  \centering
\hspace*{-1cm}
\includegraphics[scale=0.25]{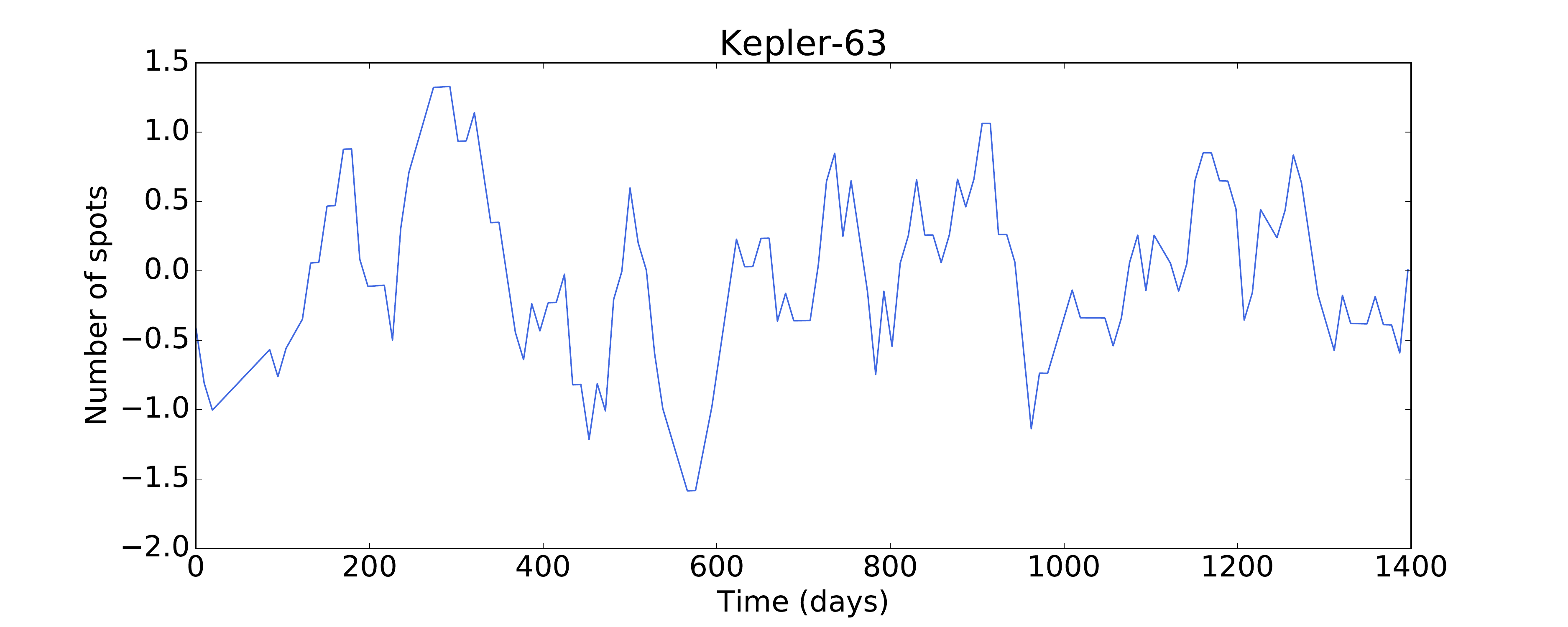}
\hspace*{-1cm}
\includegraphics[scale=0.25]{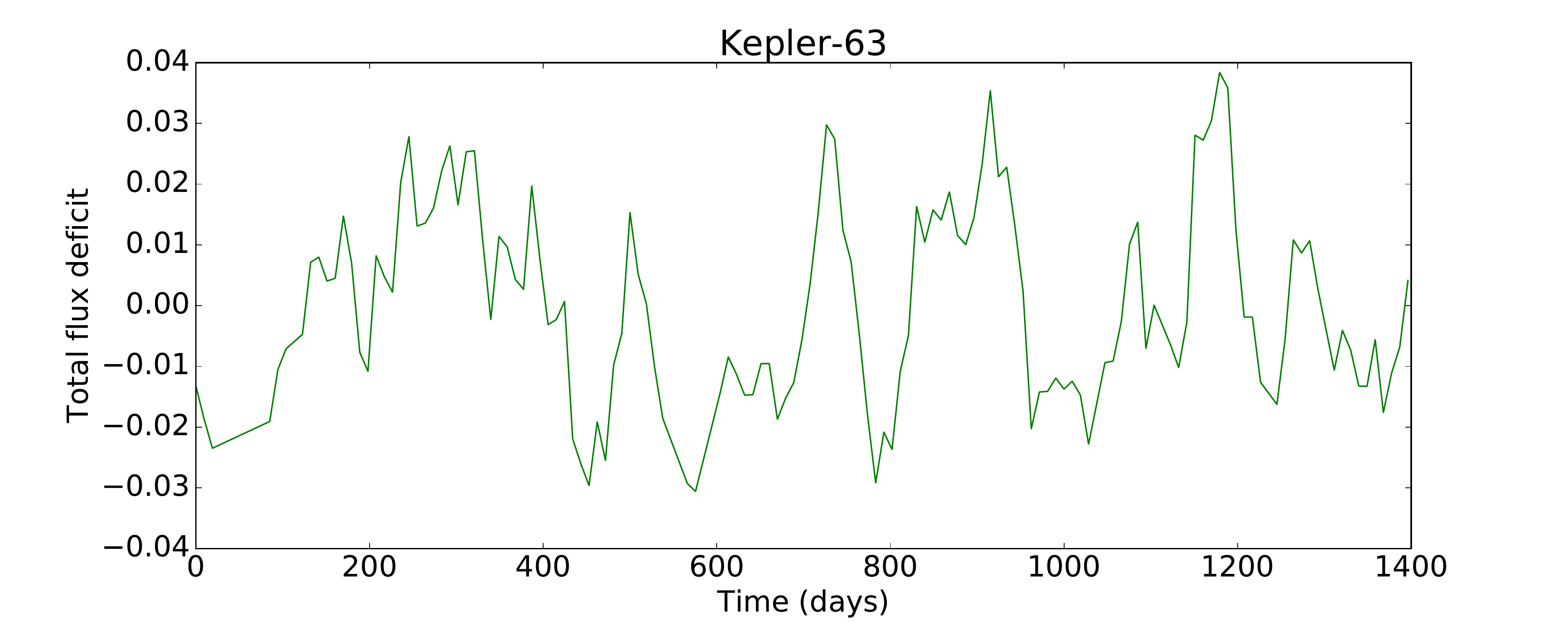}
  \caption[Flux]{Number of spots ({Top}) and total flux deficit (\textit{bottom}) in Kepler-63.}
\label{fig:figura1_6}
\end{figure}

Another way to estimate stellar activity is calculating the flux deficit resulting from the presence of spots on the star surface. The spots contrast is taken to be $1-f_{i}$, where $f_{i}$ is the relative intensity of the spot with respect to the disk center intensity $I_{c}$. A value of $f_{i}$ = 1 means that there is no spot at all. The relative flux deficit of a single spot is the product of the spot contrast and its area, thus for each transit the total flux deficit associated with spots was calculated by summing all individuals spots: 

\begin{equation}
F \approx \sum (1-f_{i})(R_{\rm spot})^{2}
\end{equation} 

To the resulting flux deficit and spot number we apply a running mean over a range of five data points. Note that due to this process, the number of spots may have non-integer values. Also, to remove possible aperiodic or long duration trends in these time series, we applied a quadratic polynomial fit to the data and then subtracted it. Due to this procedure, the number of spots and the flux deficit can be negative. Figures \ref{fig:figura1_5} and \ref{fig:figura1_6} show these treated results for the number of spots (top, in blue) obtained from modelling each transit and the total flux deficit (bottom, in green) for both stars.

\begin{figure*}[!ht]
  \centering
\includegraphics[scale=0.45]{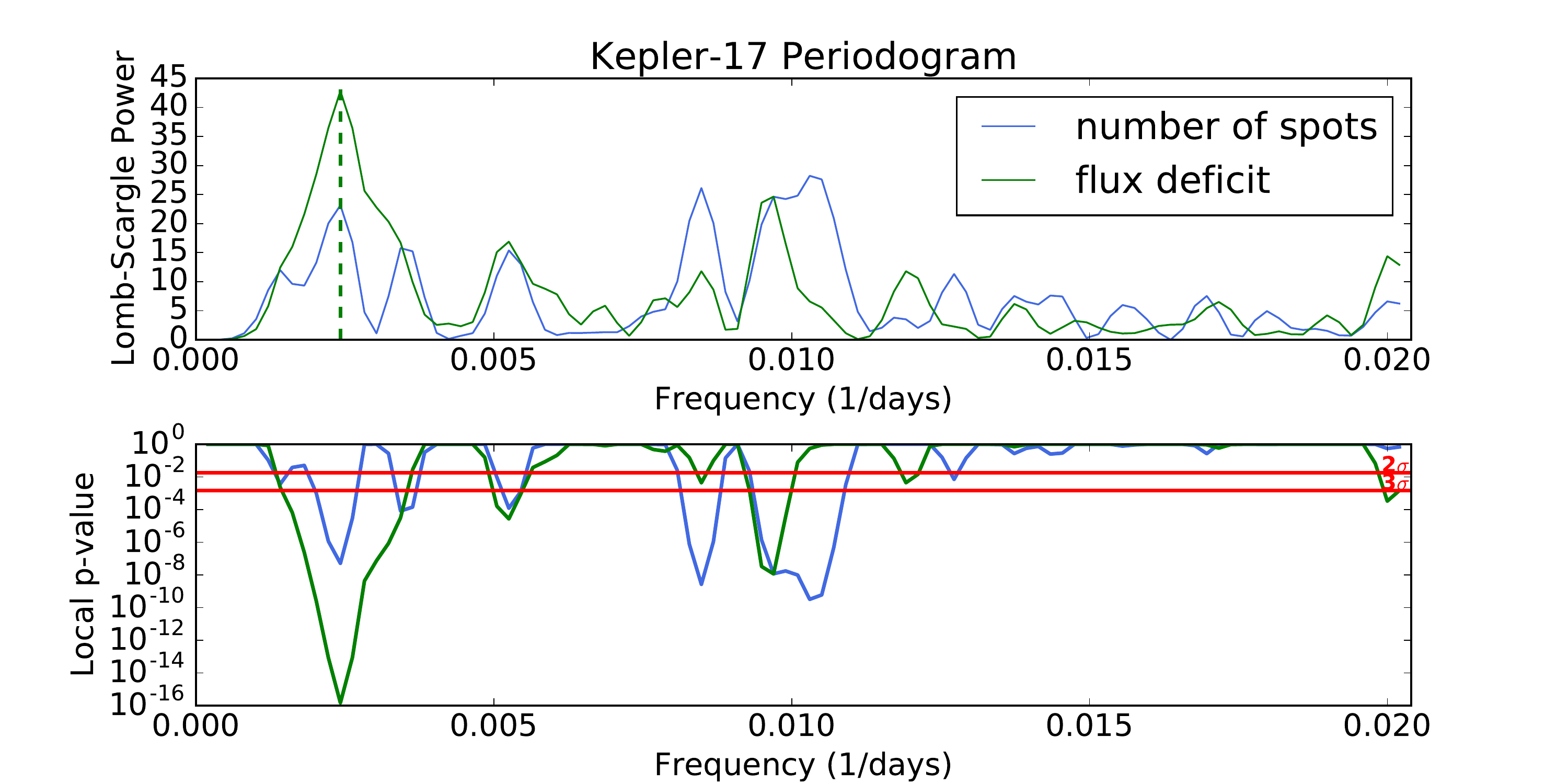}
  \caption[Lomb]{(\textit{Top}) Power spectrum indicating the periodicity of the spots number (blue) and of the flux deficit (green) for Kepler-17. The highest peak is indicated by the vertical dashed lines and is 410 $\pm$ 60 days for the spot number and 410 $\pm$ 50 days for the spot flux. (\textit{Bottom}) P-values associated to each frequency from the periodogram, the significance level is assumed to be 3$\sigma$.}
\label{fig:figura1_7}
\end{figure*}

\begin{figure*}[!ht]
  \centering
\includegraphics[scale=0.45]{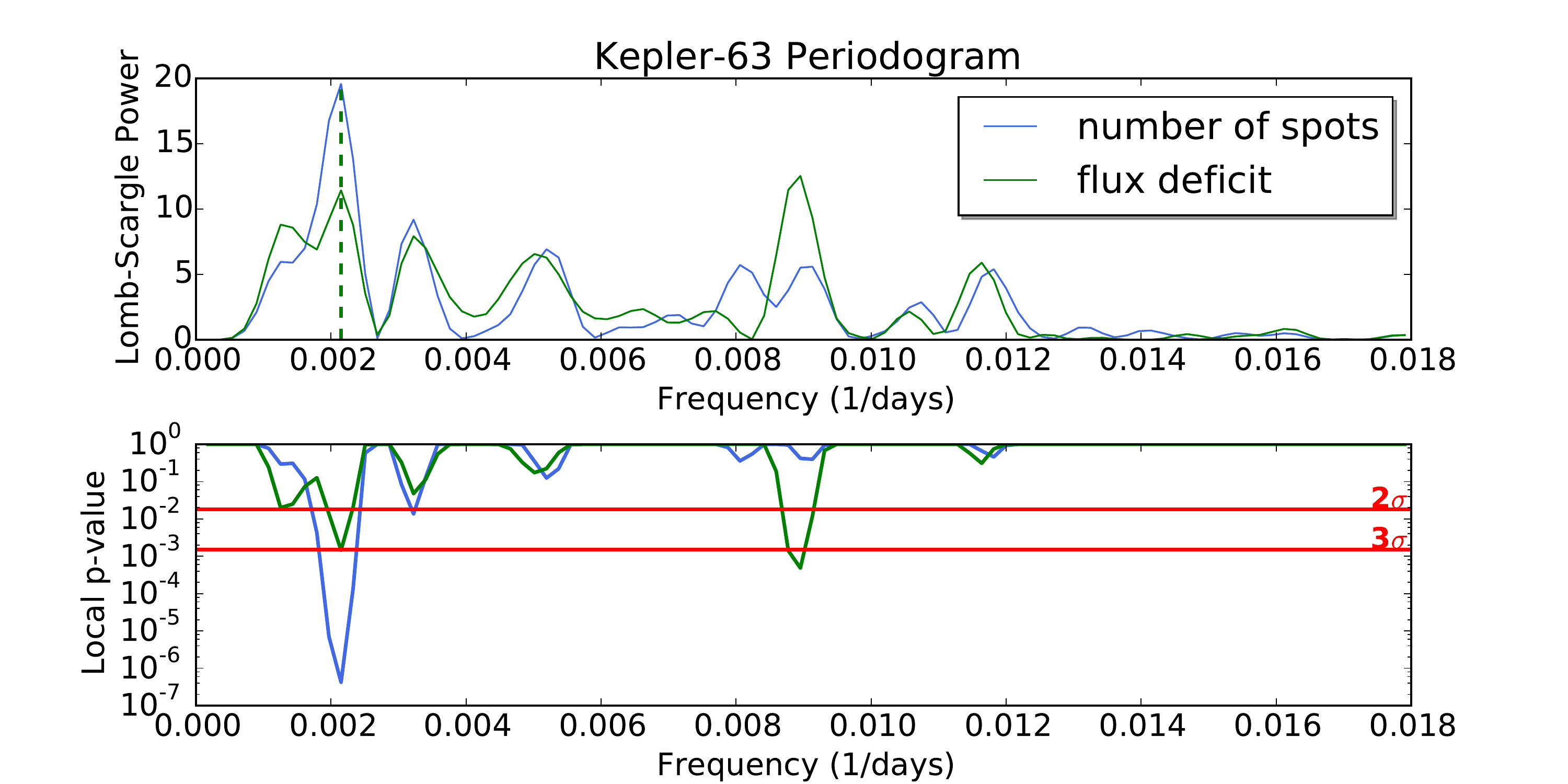}
  \caption[Lomb]{Power spectrum indicating the periodicity of the spots number (\textit{top}) and of the flux deficit (\textit{bottom}) for Kepler-63. The highest peak is indicated by the dashed lines and is 460 $\pm$ 60 days for the spot number and 460 $\pm$ 50 days for the spot flux.}
\label{fig:figura1_8}
\end{figure*}

\begin{figure*}[!ht]
  \centering
\includegraphics[scale=0.40]{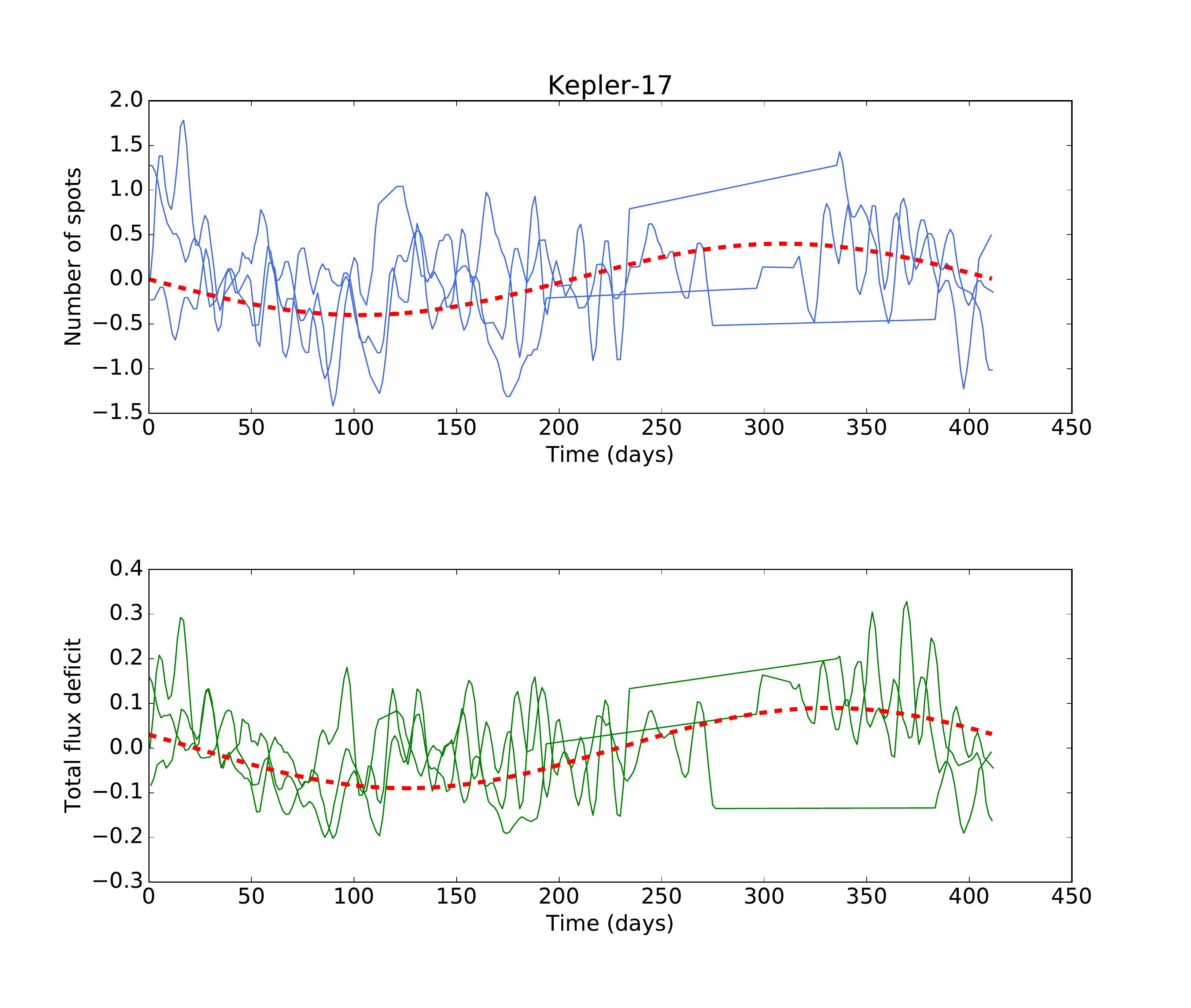}
  \caption[Folded]{Three folded data of Kepler-17 for both spot number and total flux deficit, the 410 days periodicity clearly showing.}
\label{fig:figura1_9}
\end{figure*}

\begin{figure*}[!ht]
  \centering
\includegraphics[scale=0.40]{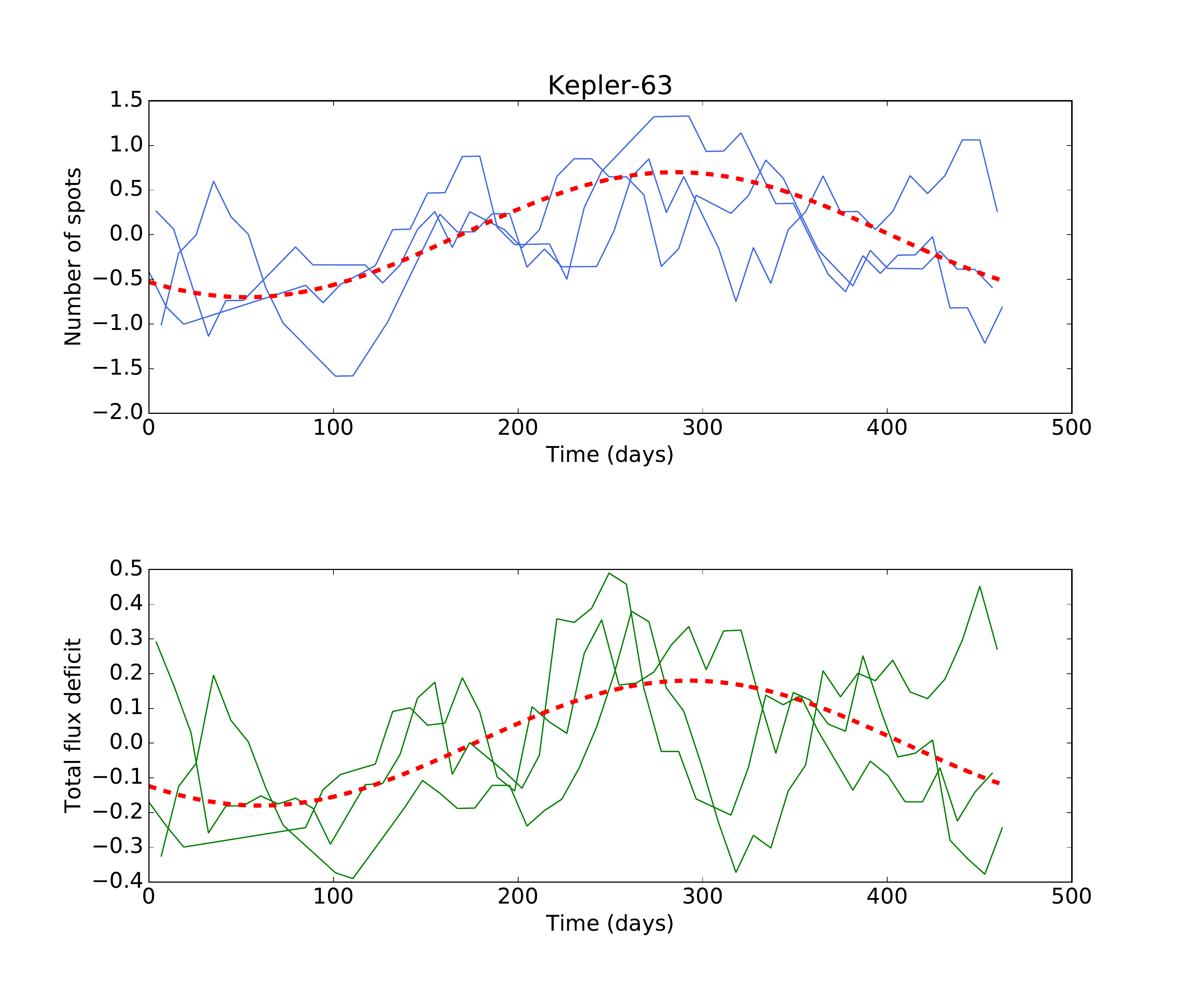}
  \caption[Folded]{Same as Fig.\ref{fig:figura1_9} for Kepler-63, indicating the periodicity of 460 days of the spot data.}
\label{fig:figura1_10}
\end{figure*}

\begin{figure*}[!ht]
  \centering
\hspace*{-7mm}
\includegraphics[scale=0.39]{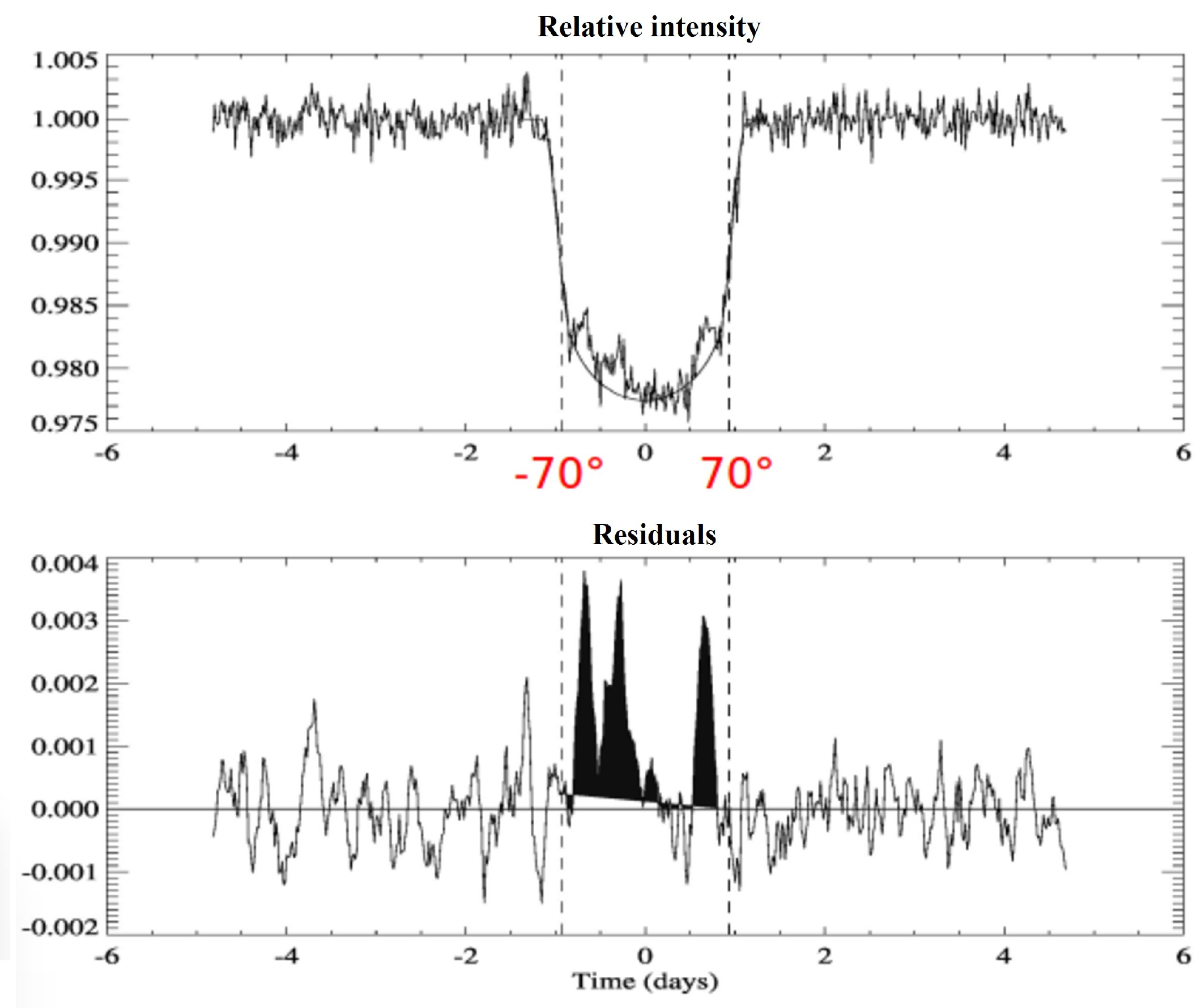}
  \caption[Subtracted transit]{\textit{Top}: 100th transit lightcurve of Kepler-17, overplotted is a simulated lightcurve without spots (solid curve). \textit{Bottom}: Residuals of the subtracted light curve. The dashed vertical lines correspond to longitudes $\pm$ 70$^{\circ}$.}
\label{fig:figura1_11}
\end{figure*}

\begin{figure}[!ht]
  \centering
\hspace*{-7mm}
\includegraphics[scale=0.24]{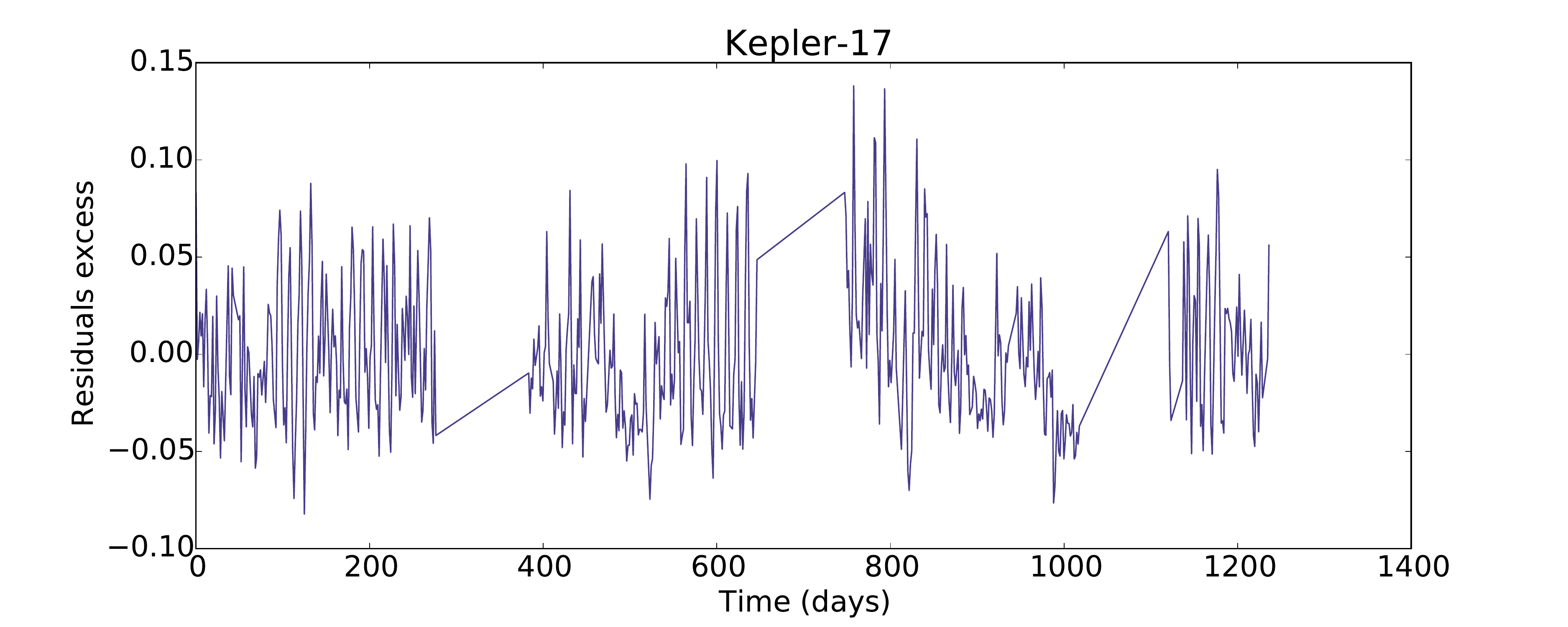}
\hspace*{-7mm}
\includegraphics[scale=0.24]{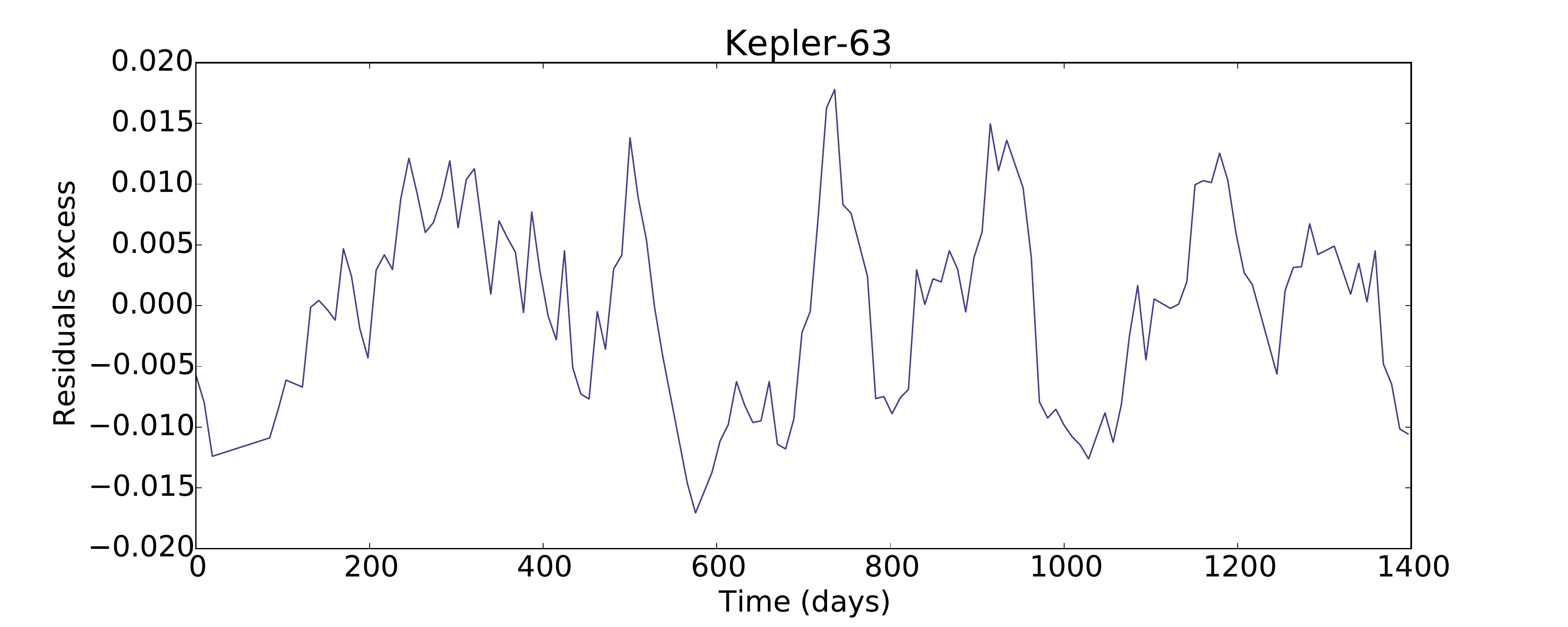}
  \caption[Activity]{Activity level from the integration of all lightcurve residuals excess during transits as described in Sect.~\ref{sec:Residuals} for Kepler-17 (\textit{top}) and Kepler-63 (\textit{bottom}).}
\label{fig:figura1_12}
\end{figure}

A Lomb Scargle periodogram (LS) \citep{scargle82} is perfomed on these time series  to obtain the period related to the magnetic cycle. In addition, we applied a significance test to quantify the significance of the peaks from the LS periodogram. For this, we assumed the null hypothesis as: (a) there is no periodicity in the data, and (b) the noise has a Gaussian distribution, thus the periodogram power spectrum in any frequency of the data will be exponentially distributed. The statistical significance associated to each frequency in the periodogram is determined by the p-value (\textrm{p}). The smaller the p-value, the larger the significance of the peak.  The corresponding values for p-value in critical values are given by $z = \sqrt(2) \rm erf^{-1} (1-2p)$. We adopted the significance level $\alpha$ as being 3$\sigma$ (\textrm{p} $\pm$ 0.0013). If the p-value is less or equal to the significance level, our null hypothesis is rejected, otherwise, it is not rejected. The result of the significance test is plotted below the periodogram for each star, in Figures~\ref{fig:figura1_7} and ~\ref{fig:figura1_8}. The uncertainty of the peaks in the periodogram is given by FWHM of the peak power.

By computing the Lomb-Scargle periodogram, it is possible to detect the existence of a periodicity in the spot number and flux deficit throughout the 4 years of observation. This result is presented on Figures \ref{fig:figura1_7} and \ref{fig:figura1_8}. In both situations, a long term periodicity is found. For Kepler-17, a prominent peak is seen at 410 $\pm$ 60 days (number of spots) and 410 $\pm$ 50 days (flux deficit). There is also another significant periodicity in $\sim$ 104 days ($f$ = 0.0096 days$^{-1}$) appearing in both cases. This peak might be a harmonic, because it is approximately four times the value of the frequency from the main peak ($f$ = 0.0024 days$^{-1}$, P = 410 days). However, as we are interested only in long term periodicities, we adopted $P_{\rm cycle}$ = 410 days as being the most relevant peak for our case. This yields a magnetic cycle of approximately 1.12 yr.

Kepler-63 shows a periodicity of 460 $\pm$ 60 days taking into account the total number of spots, and 460 $\pm$ 50 days for the flux deficit, as shown in Fig.~\ref{fig:figura1_8}. This corresponds to a cycle of about 1.27 $\pm$ 0.20 yr. There is also a strong peak at $\sim$ 111 days for the total flux deficit, however this period appears at the number of spots periodogram with a very small significance, thus we do not consider this peak.

We folded the data of both spot number and total flux deficit by the dominant period obtained in the periodograms. These results are presented in Figures \ref{fig:figura1_9} (Kepler-17) and \ref{fig:figura1_10} (Kepler-63), and in both cases we obtained three periods during the time of observation of the stars (1240 days for Kepler-17 and 1400 days for Kepler-63).




\subsection{Transit residuals}\label{sec:Residuals}

The second method adopted in this work consists in subtracting from the transit lightcurves a modelled lightcurve of a star without spots. The result from this subtraction is the residual. An example of this method is shown in Figure~\ref{fig:figura1_11} for the 100th transit of Kepler-17. An excess during the transit is clearly seen and is due to the presence of spots occulted by the planet. Next we integrated all residuals excess within $\pm$ 70$^{\circ}$ longitude of the star (delimited by the dashed lines of Fig.~\ref{fig:figura1_11}) thus obtaining another proxy for the stellar activity. This residual excess resulting from all transits is plotted in Fig.~\ref{fig:figura1_12}, and is seen to oscillate during the period of observation.

\begin{figure*}[!ht]
  \centering
\includegraphics[scale=0.40]{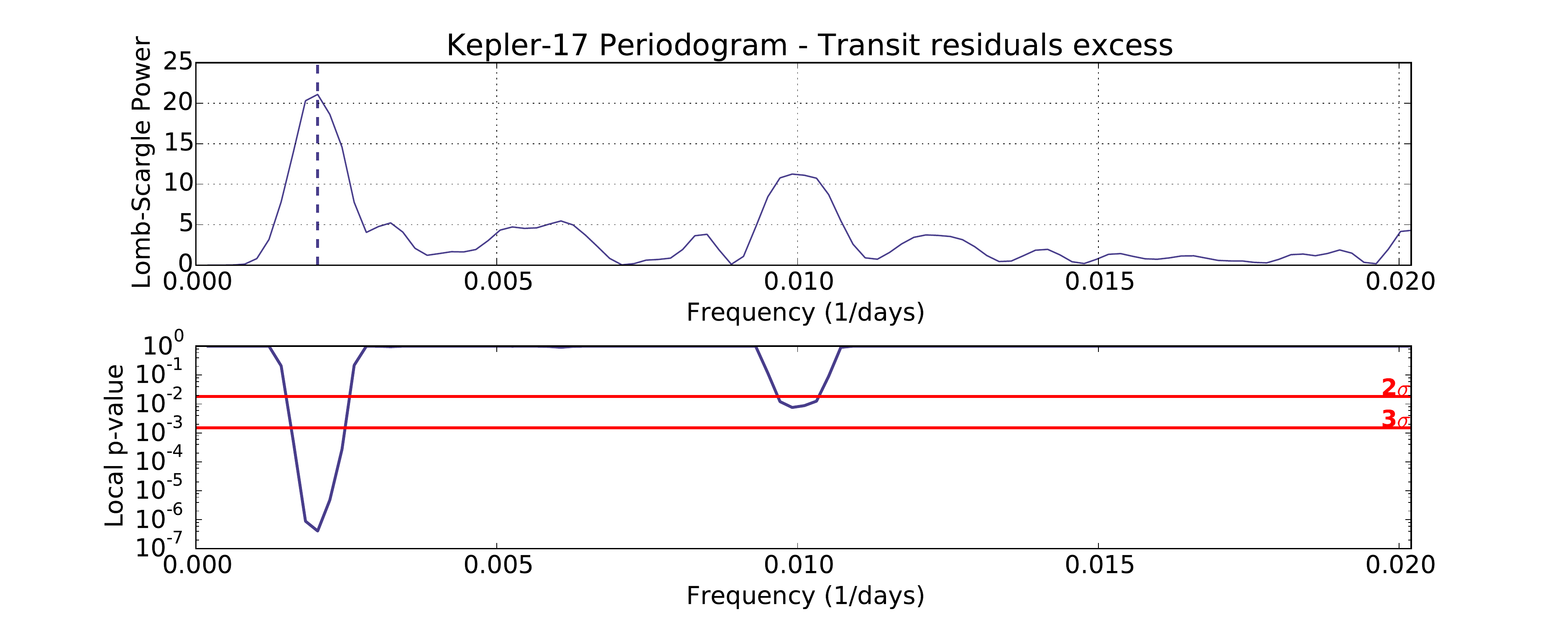}
\includegraphics[scale=0.40]{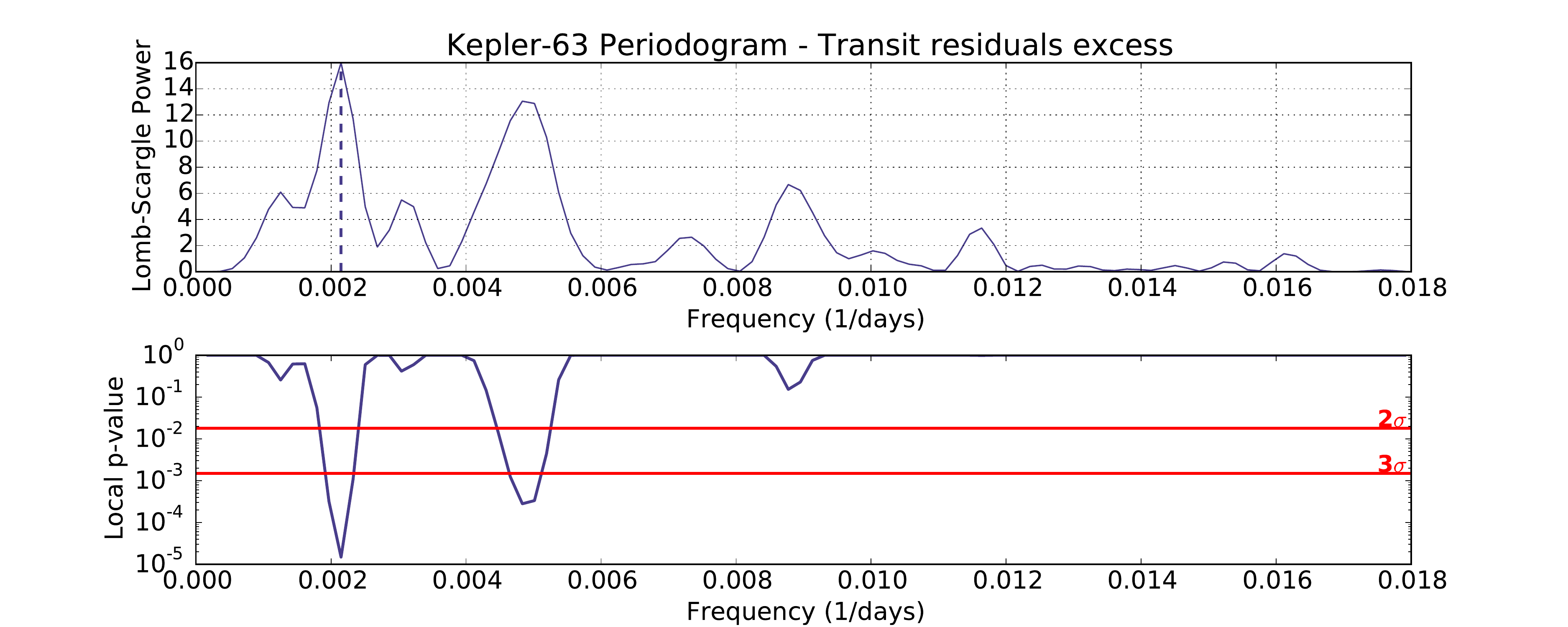}
  \caption[Lomb]{Lomb Scargle periodogram applied to the integrated transit residuals excess of Kepler-17 (\textit{top}) and Kepler-63 (\textit{bottom}). The highest peak, indicated by a dashed line, corresponds to a periodicity of 494 $\pm$ 100 days for Kepler-17 and 465 $\pm$ 40 days for Kepler-63.}
\label{fig:figura1_13}
\end{figure*}

\begin{figure*}[!ht]
  \centering
\includegraphics[scale=0.40]{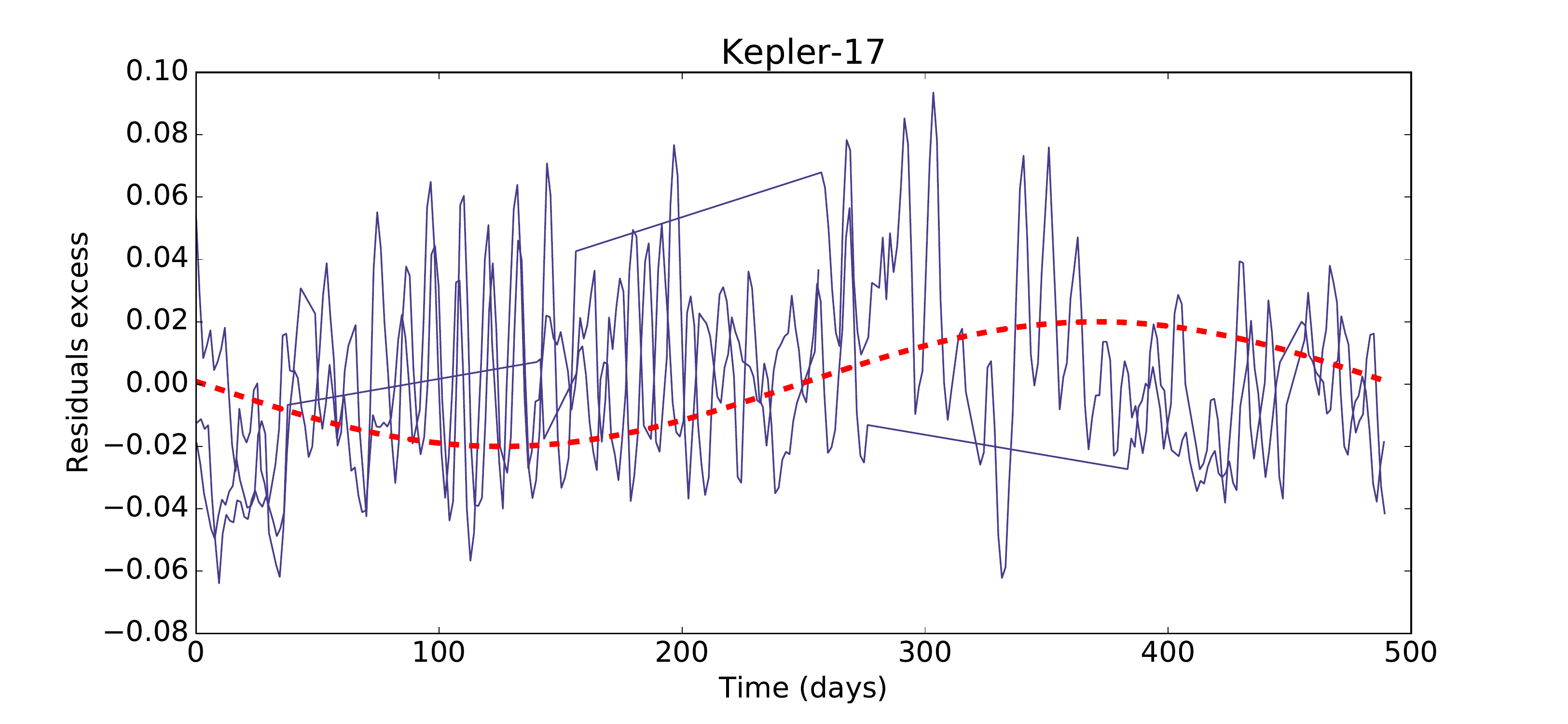}
\includegraphics[scale=0.40]{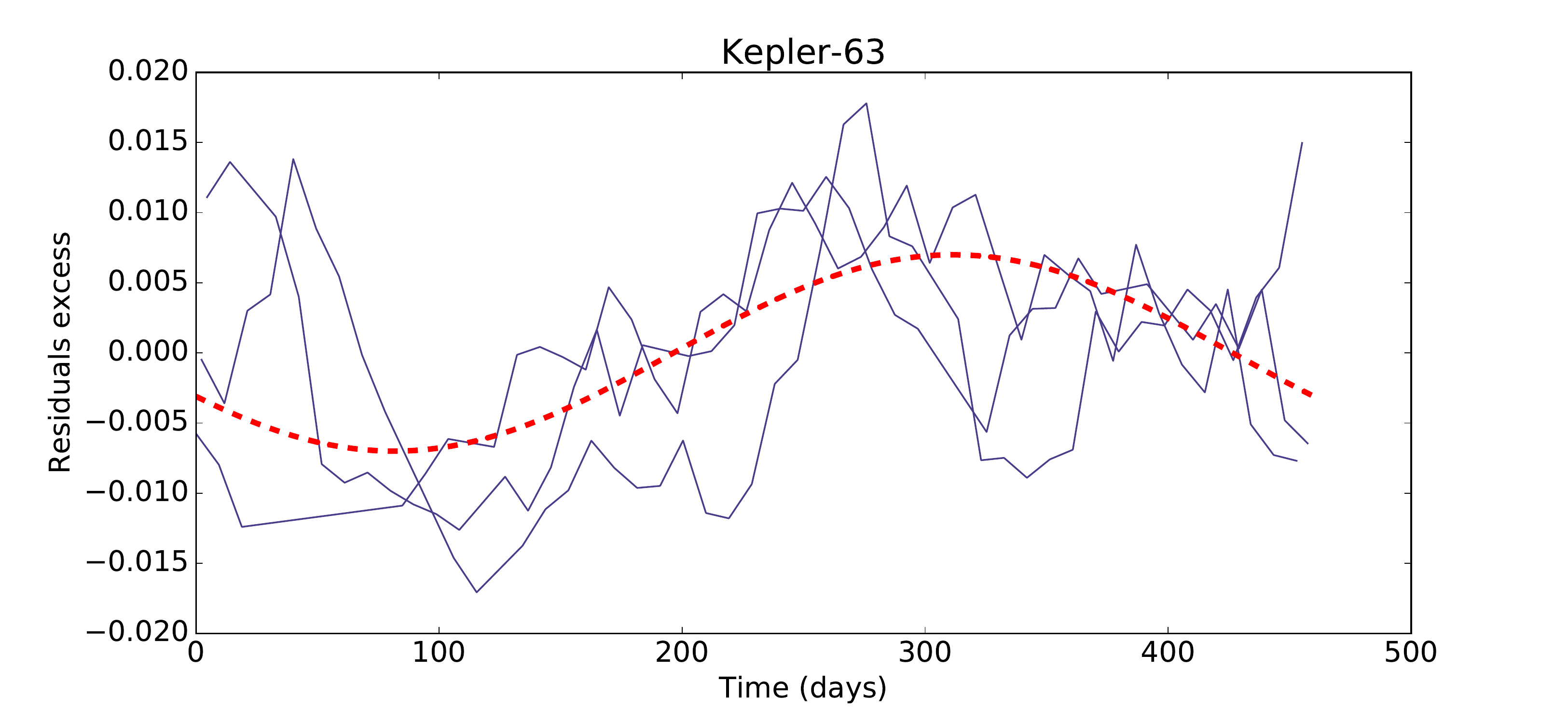}
  \caption[Folded]{Activity level data folded three times with the resulting periodicity. \textit{Top}: Kepler-17 (490 days) and \textit{bottom}: Kepler-63 (460 days).}
\label{fig:figura1_14}
\end{figure*}

\begin{figure*}[!ht]
  \centering
\includegraphics[scale=0.35]{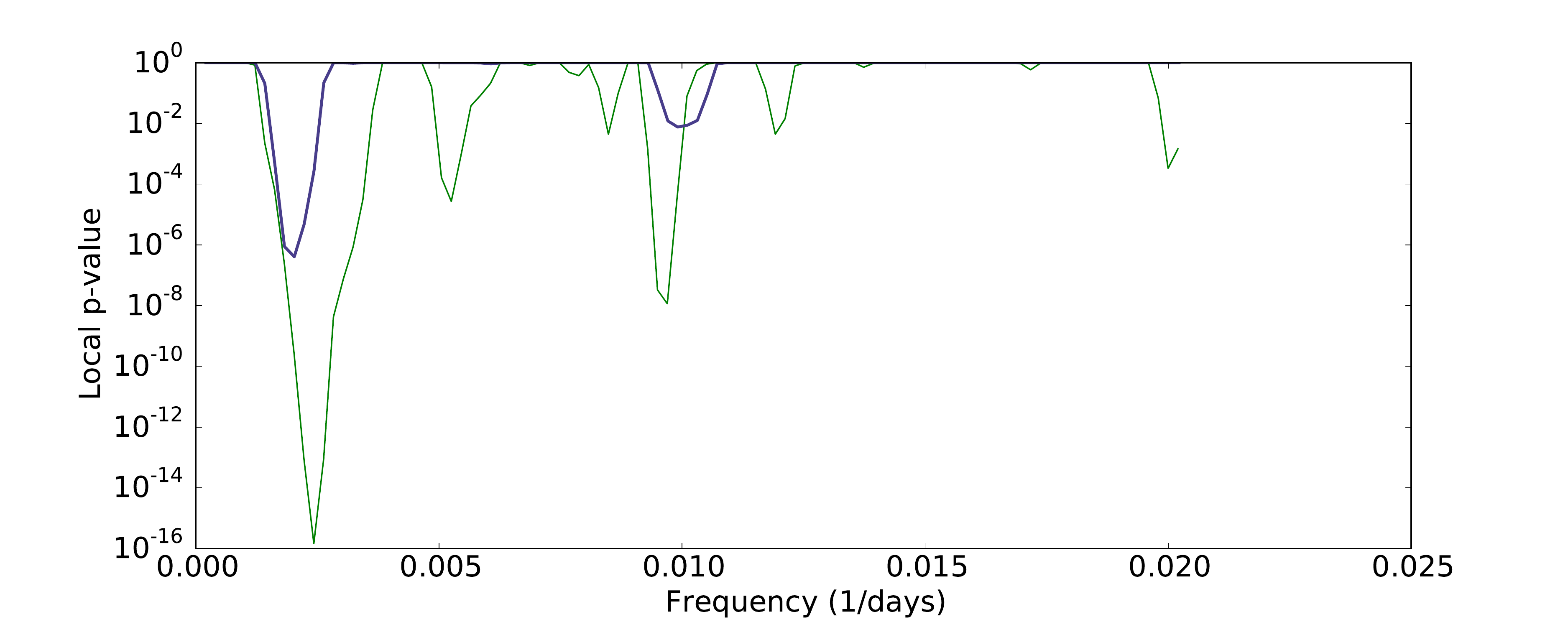}
  \caption[Significance]{Comparison between the peak significance obtained from the first method for the flux deficit (green) and the second method (purple). The frequency peak obtained in the second method falls inside the uncertainty from the first approach.}
\label{fig:figura1_15}
\end{figure*}

A quadratic polynomial fit was also applied to this time series and subtracted, removing any possible trends. Its Lomb Scargle periodogram is shown in Fig.~\ref{fig:figura1_13}, where a noticeable peak is seen at 490 $\pm$ 100 days (1.35 $\pm$ 0.27 yr) for Kepler-17 and 460 $\pm$ 40 days (1.27 $\pm$ 0.12 yr) for Kepler-63. These periodicities show a p-value below the 3$\sigma$ significance level, confirming their significance. The value obtained for Kepler-63 is similar to that from the first approach, and corresponds to a 1.27 year-cycle. There is also a notable peak at $\sim$ 205 days ($f$ = 0.00485 days$^{-1}$), but this peak does not appear in the first approach (neither for spot number or flux deficit), thus we do not consider it. The three folded data with the obtained periodicity is shown in Figure \ref{fig:figura1_14} for Kepler-17 and Kepler-63. On the other hand, the cycle period estimate for Kepler-17 agree within the uncertainty, as shown in Figure \ref{fig:figura1_15}. All the results for the magnetic cycle obtained with the two methods are listed in Table \ref{table:tab4}.

\begin{table*}[!ht]
\refstepcounter{table}\label{table:tab4}
\resizebox{1.0\textwidth}{!}{\begin{minipage}{\textwidth}
\centering
\begin{tabular}{lccc}
\multicolumn{4}{c}{\textbf{Table 4}}                                      \\
\multicolumn{4}{c}{Magnetic activity cycle periods}                       \\
\toprule
\toprule
          & \multicolumn{2}{c}{Spot Modelling} & Transit residuals excess \\
          & (Number of spots)   & (Spot Flux)  &                          \\
\midrule
Kepler-17 & 1.12 $\pm$ 0.13 yr    & 1.12 $\pm$ 0.16 yr      & 1.35 $\pm$ 0.27 yr                  \\
Kepler-63 & 1.27 $\pm$ 0.16 yr  & 1.27 $\pm$ 0.14 yr                 & 1.27 $\pm$ 0.12 yr                  \\
\bottomrule
\end{tabular}
\end{minipage}}
\end{table*}

\section{Discussion}
\label{sec:disc}

We have estimated the period of the magnetic cycle, $P_{\rm cycle}$, for two active solar-type stars, Kepler-17 and Kepler-63, by applying two new methods: spot modelling and transit residuals excess. The results obtained from both methods are presented in Table \ref{table:tab4}. For Kepler-63, we found the same result in the different approaches. The other star, Kepler-17, however, has a $P_{\rm cycle}$ obtained with the second method that is within the uncertainty range from the first approach. Considering that, we confirm that the results in both methods agree with each other and assure the robustness of the methods.
Next, we compare our results to those stars, with their magnetic cycle, $P_{\rm cycle}$, reported in the literature. The observational data from \cite{saar99} and \cite{lor05} are the records of the CaII H$\&$K emission fluxes of the stars observed at the Mount Wilson Observatory and also analysed by \cite{bau95}. \cite{olah09} studied multi-decadal variability in a sample of active stars with photometric and spectroscopic data observed during several decades. \cite{mg02} performed long-term photometric monitoring of solar analogues. \cite{lovis11} took a sample of solar-type stars observed by the HARPS\footnote{High Accuracy Radial velocity Planet Searcher} spectrograph and used the magnitude and timescale of the Ca II H$\&$K variability to identify activity cycles.

\begin{figure*}[!ht]
  \centering
\includegraphics[scale=0.35]{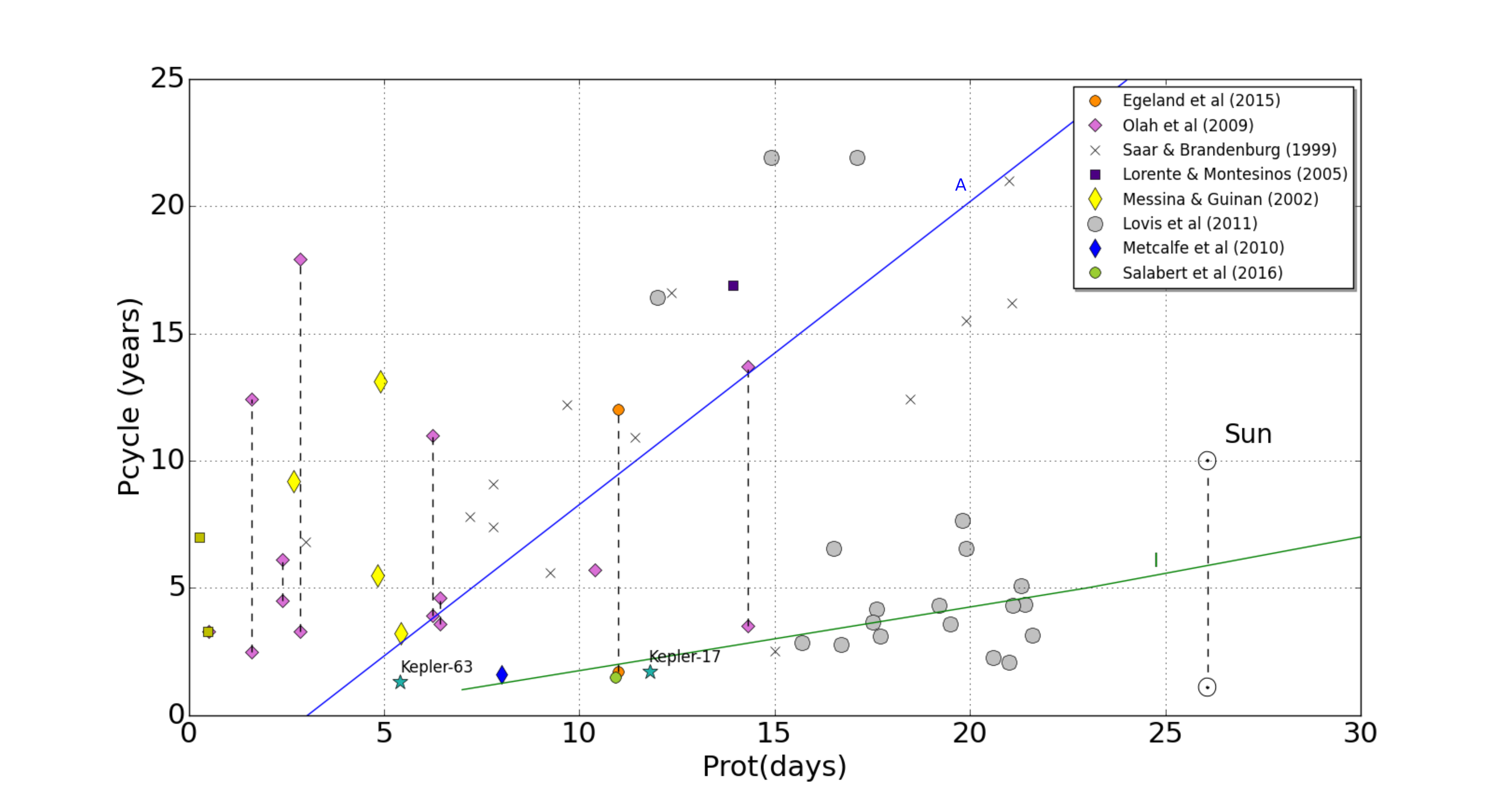}
  \caption[Cycle]{Activity cycle periods, $P_{\rm cycle}$, versus stellar rotation period, $P_{\rm rot}$, for a sample of stars cited in the literature. The dashed vertical lines connect different periods found for the same stars, i.e: stars exhibiting two cycles.}
\label{fig:figura1_16}
\end{figure*}

Two distinct branches were reported for the cycling stars \citep{saar99, bohm07}: the active (blue line) and inactive (green line), classified according to their activity level and number of rotations per cycle (plotted in Fig. \ref{fig:figura1_16}). The majority of  the stars analysed by \cite{lovis11} falls in the inactive branch, the Sun, however, with its 11 year-cycle appears in between the two branches. In Figure \ref{fig:figura1_16} we observe the relation between $P_{\rm cycle}$ (stellar cycle period) and $P_{\rm rot}$ (stellar rotation period) for all selected samples, first studied by \cite{bau96}. The vertical dashed line joins data for the same star (stars with multiple periodicities detected). Kepler-63 (blue star in  Fig. \ref{fig:figura1_16}) is an active star and in the rotational period-cycle relation follows the trend set by stars in the active branch. On the other hand, Kepler-17 (blue star) lies close to the inactive stars branch, but shows a cycle period quite close to the short-period cycle obtained by \cite{ege15} for the active star HD 30495, of $P_{\rm cycle} \sim$ 1.7 year, $P_{\rm rot}$ = 11 days, and the one analysed by \cite{salabert16} for the young solar analog KIC 10644253 ($P_{\rm cycle}$ $\sim$ 1.5 year, $P_{\rm rot}$ = 10.91 days).

\section{Conclusions}
\label{sec:conclu}

In the present work we applied two new methods to investigate the existence of a magnetic cycle: spot modelling and transit residuals excess. Two active solar-type stars were analysed, Kepler-17 ($P_{\rm rot}$ = 11.89 days and age $<$ 1.8 Gyr) and Kepler-63 ($P_{\rm rot}$ = 5.40 days and age = 0.2 Gyr). With the first method, we obtained $P_{\rm cycle}$ = 1.12 $\pm$ 0.16 yr (Kepler-17) and $P_{\rm cycle}$ = 1.27 $\pm$ 0.16 yr (Kepler-63), and for the second approach: $P_{\rm cycle}$ = 1.35 $\pm$ 0.27 yr (Kepler-17) and $P_{\rm cycle}$ = 1.27 $\pm$ 0.12 yr (Kepler-63). The results from both methods agrees with each other, considering that the value found for Kepler-17 in the second approach falls within the uncertainty of the first method. Kepler-17 and the solar analogue HD 30495 found by \cite{ege15} have a rotation period of approximately 11 days, and both show a short-cycle, which is $\sim$ 1.7 yr for HD 30495. However, this star also shows a long cycle of $\sim$ 12 years, that agrees well with the active branch. As observed previously by \cite{bohm07}, some stars that are located in the active branch, could also show short cycles falling in the inactive branch. This might be the case of Kepler-17, that is also an active star, showing a spots area coverage of 6$\%$, much higher than the Sun, where it is less than 1$\%$, and may also have a long cycle periodicity. Unfortunately, as we are constrained by the period of observation ($\leq$ 4 years) of the Kepler telescope, it is not possible to investigate if Kepler-17 has a longer cycle. On the other hand, $P_{\rm cycle}$ of the young star Kepler-63, fits well within the active stars branch. Morever, \cite{meta10} found a 1.6 year magnetic cycle for the exoplanet host star $\iota$ Horologii observed during 2008 and 2009, that seems to be in between Kepler-17 and Kepler-63 in the cycle-rotation relation.

The intensity of a magnetic field is controlled by the dynamo process, and associated with  differential rotation. Our analysis to identify activity cycle periods is essential for a deep investigation about how stellar dynamos work. \cite{bohm07} and \cite{durn81} suggest that the two branches of stars in the $P_{\rm rot}-P_{\rm cycle}$ diagram are possibly ruled by different kinds of dynamo, exhibiting different ratios for $P_{\rm cycle}/P_{\rm rot}$. In addition, B\"ohm-Vitense brings up that the active branch can be driven by a dynamo operating in the near-surface shear layer, while in the inactive stars branch, the shear layer of the dynamo is located at the base of the convection zone. This combination between the analysis of the time variation in the stellar activity and the stellar rotation periods can also be crucial to determine the differential rotation rates of the star, which is fundamental to generate the magnetic field in the stellar interiors. All these indicate that activity cycles play a key role in understanding the evolution of stars.

\acknowledgments

This work has been supported by grant from the Brazilian agency FAPESP ($\#$2013/10559-5). Raissa Estrela acknowledges a CAPES fellowship.





\appendix

\bibliography{bibliography}{}
\bibliographystyle{apalike}
\end{document}